\documentclass[preprint,12pt,sort&compress]{elsarticle}
\usepackage{amssymb}

\usepackage{amsmath}
\usepackage{algorithm}
\usepackage{algorithmic}
\usepackage{eurosym}
\usepackage{multirow}
\usepackage{pifont} 
\usepackage{array}
\usepackage{booktabs}
\usepackage{orcidlink}
\newcolumntype{P}[1]{>{\centering\arraybackslash}p{#1}}

\journal{Sustainable Energy, Grids and Networks}

\begin{document}

\begin{frontmatter}



\title{Dynamic and Risk-Aware Procurement of Fast Frequency Response from Electric Vehicles and Heat Pumps}

\author[label1,label2]{Savvas Panagi\,\orcidlink{0009-0008-6057-3392}}
\author[label3,label4]{Fanourios Kourougianni\,\orcidlink{0009-0004-2209-8923}}
\author[label3]{Andreas V. Olympios\,\orcidlink{0000-0002-5795-0408}}
\author[label1]{Chrysovalantis Spanias\,\orcidlink{0000-0003-3046-3287}}
\author[label2]{Petros Aristidou\,\orcidlink{0000-0003-4429-0225}}

\affiliation[label1]{organization={Distribution System Operator, Electricity Authority of Cyprus},
            city={Nicosia},
            country={Cyprus}}
\affiliation[label2]{organization={Department of Electrical and Computer Engineering and Informatics, Cyprus University of Technology},
            city={Limassol},
            country={Cyprus}}
\affiliation[label3]{organization={PHAETHON Centre of Excellence for Intelligent, Efficient and Sustainable Energy Solutions and the PV Technology Laboratory, Department of Electrical and Computer Engineering, University of Cyprus},
            city={Nicosia},
            country={Cyprus}}
\affiliation[label4]{organization={Department of Mechanical and Manufacturing Engineering, University of Cyprus},
            city={Nicosia},
            country={Cyprus}}

\begin{abstract}
Low-inertia power systems increasingly rely on Fast Frequency Response (FFR) to keep the frequency nadir above under-frequency load-shedding thresholds without oversized Frequency Containment Reserve or synchronous inertia. This paper proposes two complementary methodologies that interface with an existing day-ahead FFR capacity market. On the Transmission System Operator (TSO) side, a dynamic procurement rule maps forecasted kinetic energy and contingency magnitude onto an offline neighborhood-need surface, yielding interval-wise FFR capacity requests rather than constant seasonal or annual volumes. On the aggregator side, a risk-aware, TSO-certifiable bidding strategy quantifies reliable FFR capacity from fleets of electric vehicles (EVs) and heat pumps (HPs) by offering lower conditional quantiles of day-ahead headroom at a prescribed delivery reliability. The resulting requests and offers are cleared through a pay-as-bid capacity auction. A Cyprus 2025 case study shows that dynamic neighborhood-based procurement yields an annual clearing cost of about \euro476{,}000, which is 30\%-33\% lower than the seasonal and annual static rules and also below the UFLS VoLL of \euro637{,}500 estimated from the fifteen real 2025 events with $f_{\min}\le 49\,\mathrm{Hz}$. EV and HP offers meet their target delivery reliabilities, and RMS validation finds 46 of 50 operating points above 49Hz at both full and 85\% FFR delivery.
\end{abstract}




\end{frontmatter}



\section{Introduction}

\subsection{Motivation}

Maintaining a continuous balance between power generation and demand is essential for stable power system operation. Any imbalance between generation and consumption causes synchronous generators (SG) to accelerate or decelerate, resulting in frequency deviations governed by the swing equation \cite{Future-System-Inertia}. Such imbalances are typically caused by unexpected generation or load disturbances and can lead to severe frequency excursions. However, the ongoing transition towards renewable energy sources (RES), driven by sustainability goals \cite{IEA2023NetZero}, has led to the progressive decommissioning of conventional SG, significantly reducing both system inertia and the primary frequency response traditionally provided by these units. Consequently, low-inertia power systems have become increasingly vulnerable to frequency instability.

Studies have shown that, even when the Rate of Change of Frequency (RoCoF) and steady-state frequency remain within their limits (due to sufficient inertia and primary frequency reserve), the frequency nadir may still violate security requirements, triggering under-frequency load shedding (UFLS) \cite{iet_savvas_impact_ffr}. To address this limitation without increasing the minimum inertia or Frequency Containment Reserve (FCR) requirements, the Nordic Transmission System Operators (TSOs) introduced Fast Frequency Response (FFR) (see Fig. \ref{fig:ffr_activation_example}). FFR has proven to be an effective and cost-efficient solution, as it improves the frequency nadir without requiring RES curtailment or additional FCR procurement \cite{statnett_ffr}. Motivated by this, from the provider's perspective, Distribution Network (DN)-connected resources, such as Electric Vehicles (EVs) and Heat Pumps (HPs), can participate in FFR markets through aggregators, creating an additional revenue stream for asset owners.

\subsection{Literature Review and Research Gap}

A substantial part of the literature models FFR through droop-based control \cite{VATSAL2025111778}, rather than as the short-duration, step-like response introduced in the Nordic FFR framework (see Sec. \ref{sec:background}). Such approaches are more closely related to FCR, in which power is continuously adjusted according to the frequency deviation. In contrast, Nordic FFR is a distinct threshold-activated product introduced to improve frequency security in low-inertia conditions while avoiding the higher operational costs associated with additional FCR or synchronous inertia. This distinction changes how aggregators quantify and offer capacity and how TSOs determine and procure the required reserve, an aspect that has received limited attention in the literature. Step-like FFR from DN-connected resources has mainly been examined at the device level; for instance, a commercial home charger was modified in \cite{manner2024ev_ffr} to meet the Nordic FFR activation requirements in a laboratory proof of concept, without an aggregator-level quantification of the deliverable capacity.

An hourly security-constrained unit commitment (SCUC) model that co-optimizes energy and frequency reserves is proposed in \cite{KUSHWAHA2023106944}, jointly determining conventional generation and EV schedules, FFR and primary frequency reserves, and EV charging. A similar SCUC framework with additional provision components is presented in \cite{VATSAL2025111778}. Although these approaches enable the coordinated scheduling of energy and frequency services, their large-scale formulations are computationally intensive and require extensive exchange of resource-level information, raising scalability and privacy concerns. More importantly, they introduce a highly integrated co-optimization structure that differs substantially from existing reserve-market arrangements, where FFR and FCR are procured as distinct products with different technical requirements and activation mechanisms (see Sec. \ref{sec:background}). Consequently, their practical adoption would require considerable changes to existing market structures rather than enabling participation within established FFR products.

Beyond the aforementioned market-structure limitations, comparatively limited attention has been given to FFR provision from thermostatically controlled loads (TCL), particularly HPs. Nevertheless, some recent studies have started to examine their potential. In \cite{song2024power}, the authors developed a power-hardware-in-the-loop platform to emulate the nonlinear transient behavior of an air-source heat pump (ASHP) under rapid power modulation. The proposed multiphysics model integrates the electrical and thermal subsystems and operates in real time, enabling the development and testing of FFR-compatible controllers. However, the study primarily considers load increases and does not demonstrate FFR provision through load reductions during under-frequency events. In a subsequent study \cite{song2024integrating}, the same authors proposed a nonlinear physics-based ASHP model with a reverse-droop controller, extending the framework to both load increases and reductions and demonstrating the technical feasibility of HP-based FFR.

From the TSO perspective, determining the required FFR capacity for each market period is a prerequisite for establishing an independently cleared FFR market. The required capacity, however, strongly depends on the prevailing system operating conditions \cite{fingrid_ffr}. This dependency has been considered in the literature, for example, in \cite{7115982}, where inertia-dependent FFR requirements are incorporated within stochastic unit commitment, and in \cite{9470941}, where FFR and primary-frequency-response requirements are jointly determined as a function of system inertia. Nevertheless, these approaches embed FFR within broader reserve co-optimization frameworks rather than deriving a standalone FFR capacity requirement before clearing an independent reserve market.

The need for time-varying FFR procurement is already recognized in Nordic operational practice \cite{fingrid_ffr,statnett_ffr}. Fingrid determines the FFR procurement need according to the prevailing system inertia and procures FFR only for hours in which it is required \cite{fingrid_ffr}, while Statnett combines a seasonal FFR profile with a flexible volume requested by system operation according to a short-term forecast of the FFR need \cite{statnett_ffr} . Similar dynamic dimensioning principles are also applied to other frequency reserves, such as Frequency Restoration Reserve (FRR) \cite{DEVOS2019272}, for which reserve requirements are determined prior to market clearing. These practices motivate the development of a systematic methodology for translating forecasted system conditions into an explicit FFR capacity requirement for each upcoming market period, which can subsequently serve as the demand input to an independently cleared FFR capacity market.

\subsection{Paper Contributions}

Considering the above, existing approaches do not provide an integrated framework that aligns with the structure and requirements of current FFR market products. To address these limitations, this work proposes an industry-oriented framework addressing both TSOs seeking to dynamically procure FFR and aggregators seeking to reliably quantify the capacity available from uncertain Distributed Energy Resources (DERs). The framework is further inspired by established reserve-market practices for stochastic resources, such as the P90 requirement adopted in the Danish reserve market \cite{10960637}, supporting the practical feasibility of the proposed approach. The main contributions are as follows:

\begin{itemize}
    \item A TSO-side pipeline that constructs an offline DSA-based neighborhood FFR need surface and turns day-ahead kinetic-energy forecasts and contingency magnitudes into interval-wise capacity requests by surface look-up, so the TSO can vary procured FFR with system conditions without re-running incident simulations at market time.
    \item A risk-aware aggregator-side FFR capability quantification for heterogeneous DER portfolios (electric vehicles and heat pumps), in which day-ahead capacity offers are formed as lower conditional quantiles of portfolio headroom under a prescribed delivery reliability, consistent with established reserve practices for stochastic resources.
    \item An end-to-end Cyprus case study that closes the loop from dynamic requests and Quantile Regression Forest (QRF) offers through pay-as-bid market clearing, benchmarks procurement cost against static rules and a UFLS Value of Lost Load (VoLL) reference, and validates frequency-nadir security, including sensitivity to partial FFR under-delivery, on representative 2025 operating conditions. 
\end{itemize}

The remainder of this paper is organized as follows. Section~\ref{sec:background} provides the background on FFR and the characteristics of suitable resources. Section~\ref{sec:dynamic_requests} develops the dynamic TSO-side FFR capacity request. Section~\ref{sec:method} presents the risk-aware aggregator-side capacity quantification and its interface to day-ahead FFR capacity clearing. Section~\ref{sec:case_study} describes the Cyprus case study, and Section~\ref{sec:Results_Discussions} presents the numerical results. Finally, Section~\ref{sec:conclusion} concludes the paper.

\section{Background}
\label{sec:background}

\subsection{Fast Frequency Response: Technical and Market Characteristics} 

\begin{figure}
    \centering
    \includegraphics[width=\columnwidth]{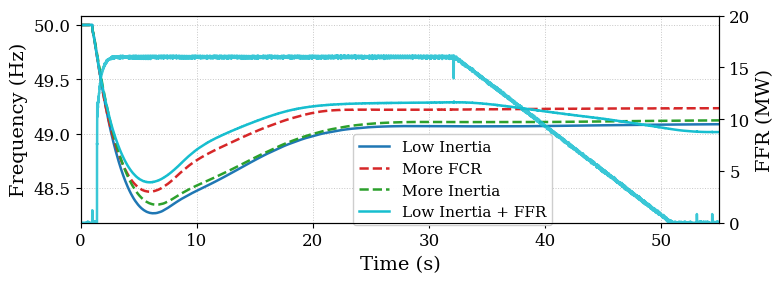}
    \caption{Frequency response after loss of generation and FFR response.}
    \label{fig:ffr_activation_example}
\end{figure}

The frequency behavior after the loss of significant generation is shown in Fig.~\ref{fig:ffr_activation_example}, with the most important parameters being the initial RoCoF, the minimum frequency (nadir), and the post-fault steady-state frequency. The limits of all three parameters are strictly defined in system grid codes, as they can significantly impact the system security. A low nadir can trigger UFLS protection (most typically at 49.0 Hz), leading to prolonged disconnection of multiple feeders and the loss of both RES generation and load consumption, both of which are costly and undesirable \cite{iet_savvas_impact_ffr}. To overcome this unwanted event, operators can increase FCR (dashed red line) or increase Kinetic Energy (KE) (dashed green line). Neither solution is desirable, as they usually require curtailment of RES generation. 

On the other hand, adding FFR (light blue line) through inverter-based resources (IBRs) can achieve the desired results without curtailing RES generation. Technically, FFR is a reserve that is activated very quickly to prevent frequency drops below a predefined threshold in the event of major faults in the power system \cite{NordicTSO2019FFR,statnett_ffr}, as depicted in Fig.~\ref{fig:ffr_activation_example}. The FFR is a step-power active response with a maximum full-activation time of 0.70, 1.00 or 1.30 s for activation thresholds of 49.5, 49.6 or 49.7 Hz, respectively, and a minimum support duration of 30 s \cite{NordicTSO2019FFR}. After the support period, the FFR controller gradually decreases the power output over the next 20 seconds. The smooth deactivation is performed to allow the FCR controllers to handle the power change. 
The total delivered energy, represented by the area under the FFR activation profile, amounts to only about $0.011\,\mathrm{kWh}$ per $\mathrm{kW}$ of FFR capacity, corresponding to approximately 1.1\% of the energy associated with one hour of operation at the reserved power, for the usual settings ($T_{\mathrm{act}}=1\,\mathrm{s}$, $T_{\mathrm{supp}}=30\,\mathrm{s}$, $T_{\mathrm{deact}}=20\,\mathrm{s}$).

From the market perspective, FFR can be procured as a reserve-capacity product, as implemented by Nordic TSOs since 2020 \cite{NordicTSO2019FFR}. FFR capacity is procured for periods specified by the system operator, and it can be divided into two products. A seasonal product is procured in limited volumes mainly during specific periods, whereas a flexible product is procured according to short-term forecasts of system needs. The latter allows the procured FFR capacity to adapt to varying operating conditions. Market clearing follows a bidding structure similar to other reserve-capacity markets. The procured volume is derived from a forecast of system inertia for the coming week, to which a transient frequency stability model is applied together with the reference incident, and it includes a margin for modeling and forecasting uncertainty \cite{NAG2025MinInertia}. 
Outside the Nordic area, Great Britain has procured a comparable sub-second post-fault response, Dynamic Containment, through day-ahead auctions since October 2020 \cite{NESO2020DC}.

\subsection{Inverter-based Resources Characteristics Favoring FFR Participation}

Owing to their power-electronic interface, IBRs are technologically capable of delivering FFR. As discussed above, FFR requires a very fast active-power response, yet the associated energy exchange is small because activations are rare and short, analogous to Nordic analyses of FCR-D provision from EV fleets \cite{10960637}. The most suitable providers are therefore IBRs that can offer upward flexibility, either by reducing demand or increasing injection, from available headroom without compromising their primary service or other market revenues. Under these criteria, this paper focuses on EVs and HPs, both of which already have an inverter that can be configured for FFR with modest controller changes, and, in residential settings, typically exceed approximately 5-7\,kW per resource, yielding a meaningful contribution when aggregated. Remaining loads are generally neither inverter-interfaced for straightforward FFR control nor large enough in rating to justify participation.

EVs clearly illustrate this suitability due to their negligible impact on battery state of charge (SoC) and high rated power. In addition, vehicles are typically grid-connected when not in use; FFR participation can therefore generate an ancillary revenue stream for owners, with limited interference with mobility needs. This additional value can partially offset charging infrastructure and battery CAPEX, further support EV adoption, and contribute to frequency support that facilitates higher renewable penetration. More generally, stationary battery storage shares the same attributes, with the practical advantage of near-continuous availability.

Thermostatically controlled loads are likewise well-suited to FFR, as the inherent thermal inertia of buildings enables short-term reductions in electrical demand while limiting deviations from acceptable indoor temperature conditions \cite{Callaway2009,Nema2025}. Heat pumps can provide such flexibility by temporarily interrupting compressor operation. Various modern residential inverter-driven split units can support externally commanded operation/stop or forced-stop functions through dedicated control interfaces (e.g., \cite{FujitsuKMCE2022,DaikinFTXMA2024}). These interfaces can be actuated by an external relay driven by a local frequency-responsive controller. Experimental studies have demonstrated the effectiveness of direct load control for heat pump demand response and frequency regulation \cite{Kim2016,Crawley2023}. Given the stringent Nordic FFR activation times of 0.7--1.3~s \cite{NordicTSO2019FFR}, locally triggered compressor shutdown offers a promising approach, as it avoids the communication delays associated with centralized dispatch \cite{Nema2025,Crawley2023}. 
Beyond these implementation considerations, a key challenge for market participation remains the reliable quantification of available flexibility, which depends on weather conditions, occupancy, building characteristics, and control settings \cite{Nema2025}, motivating the risk-aware capability assessment developed later in this work.

\section{Fast Frequency Response Dynamic Requests}
\label{sec:dynamic_requests}

This section develops the TSO-side \textit{dynamic capacity request} in the proposed FFR market framework, which determines how much FFR capacity to procure for each market interval of the next day. Because FFR needs vary strongly with system KE, contingency magnitude, and available FCR, high-inertia and high-FCR hours often require little or no FFR to keep the nadir above the UFLS threshold, and vice versa. A constant seasonal or annual maximum is therefore neither economically efficient nor operationally desirable: unnecessary reservation payments are incurred in hours where FFR is not needed, an \textit{economic inefficiency} quantified in Section~\ref{sec:results_clearing}, and an oversized step response activated on milder imbalances can overcorrect the frequency trajectory, which is the operational risk of over-procurement.

\subsection{Offline FFR Need Surface from Historical Sizing}
\label{sec:ffr_need_surface}
In~\cite{PANAGI2025101699}, FFR was sized for known operating scenarios using Dynamic Security Assessment (DSA) coupled with a gradient-descent linear estimator, yielding, for each scenario $s$, the minimum FFR power $P_{s}^{\mathrm{FFR}}$ required to restore the frequency nadir to a prescribed target (e.g., above the first UFLS stage). However, the actual operating point is not known at the time of capacity-market clearing. An offline system-behavior learning process is therefore proposed that uses historical operating conditions and their corresponding FFR requirements to construct a reusable mapping from system conditions to the required FFR capacity.

For each historical point $s\in\mathcal{H}$, the sizing process produces an empirical dataset of quadruples $\mathcal{S}=\{(KE_{s},\,FCR_{s},\,P_{s}^{\mathrm{cont}},\,P_{s}^{\mathrm{FFR}})\}_{s\in\mathcal{H}}$, where $KE_{s}$ is the pre-disturbance kinetic energy, $FCR_{s}$ is the available containment reserve, $P_{s}^{\mathrm{cont}}$ is the contingency magnitude, and $P_{s}^{\mathrm{FFR}}$ is the corresponding calculated FFR requirement. In general, $\mathcal{S}$ is not single-valued in $(KE,FCR,P^{\mathrm{cont}})$, as the required FFR also depends on the activation speed of the available FCR units and the load composition~\cite{PANAGI2025101699}. A secure lookup may therefore use an \emph{upper envelope} of $\mathcal{S}$ for robust operation.

\begin{figure}
    \centering
    \includegraphics[width=0.8\columnwidth]{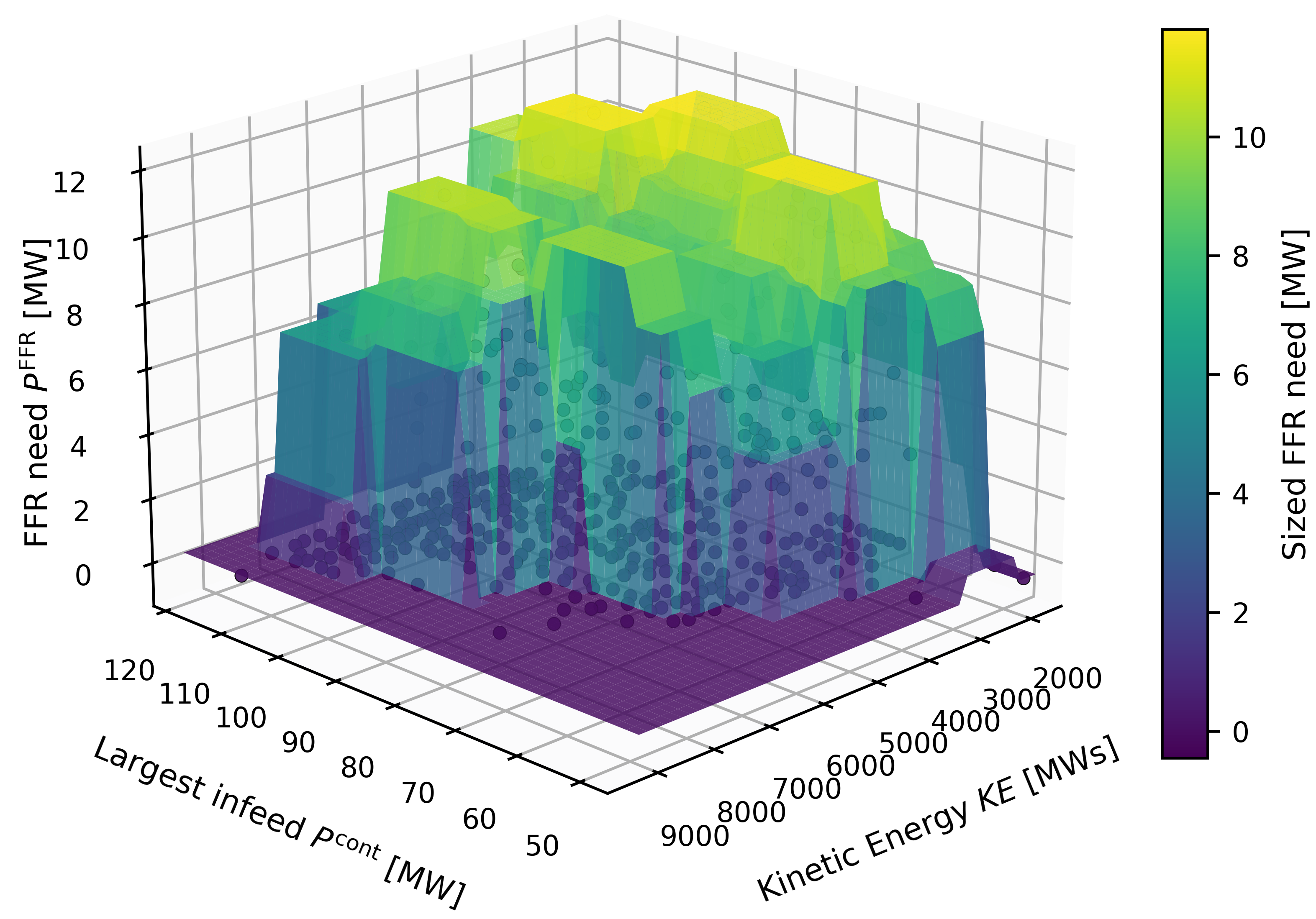}
    \caption{Offline FFR needs surface over KE and contingency magnitude. Coloured markers are sized historical scenarios; the surface is the neighborhood-wise maximum of~$\mathcal{S}$.}
    \label{fig:ffr_need_maximum}
\end{figure}

Depending on the country or region of application, the FCR procurement requirement may be static and known in advance. For instance, in Continental Europe and the Nordic synchronous area, the FCR and FCR-N requirements are defined as 3000 and 600~MW, respectively~\cite{ENTSOE_CE_SAFA_2024,NordicSOA_FCR_2025}. Therefore, the space can be reduced to triples\footnote{In the case of non-constant FCR, the methodology can be extended by including one additional parameter in the space without modifying the process.}. 
The resulting surface is obtained by a neighborhood-wise maximum over~$\mathcal{S}$. Writing each sized triple as $z_{s}=(KE_{s},P^{\mathrm{cont}}_{s},P_{s}^{\mathrm{FFR}})$, for a query $(KE,P^{\mathrm{cont}})$ the local set~$\mathcal{N}$ of historical points within tolerances $(\Delta_{\mathrm{KE}},\Delta_{\mathrm{cont}})$ is collected~\eqref{eq:ffr_nbhd}. The secure lookup~\eqref{eq:ffr_need_surface} is then the maximum sized FFR in that neighborhood, as illustrated in Fig.~\ref{fig:ffr_need_maximum}\footnote{A convex-hull upper envelope of~$\mathcal{S}$ remains attractive when the need surface must enter an optimization as a piecewise-linear constraint, e.g.\ co-optimised FFR procurement in the imbalance settlement period (ISP); this embedding is left for future work.}. 
The surface is computed once offline and can be refreshed periodically as the generation mix evolves.
\begin{align}
    &\mathcal{N}(KE,P^{\mathrm{cont}})
    =\bigl\{s\in\mathcal{H}:|KE_s-KE|\le\Delta_{\mathrm{KE}}, \nonumber\\
    &\qquad |P_s^{\mathrm{cont}}-P^{\mathrm{cont}}|
    \le\Delta_{\mathrm{cont}}\bigr\},
    \label{eq:ffr_nbhd}
    \\
    &\widehat{P}^{\mathrm{FFR}}(KE,P^{\mathrm{cont}})\!=\!\max\bigl\{P_{s}^{\mathrm{FFR}}:s\in\mathcal{N}(KE,P^{\mathrm{cont}})\bigr\}
    \label{eq:ffr_need_surface}
\end{align}

\subsection{Dynamic FFR Capacity Request}
\label{sec:ke_fcr_forecast}

For day-ahead operation, the TSO does not yet know the exact pre-disturbance operating point of each interval.
In practice, several TSOs already operate internal inertia or KE forecasting tools, and more detailed day-ahead forecasting methods have been proposed in the literature~\cite{heylen2022inertia,riquelme2024ke}. The short-term inertia prediction approach of~\cite{zhang2023xgboost}, which is based on extreme gradient boosting (XGBoost), is adopted here. In brief, XGBoost builds an additive ensemble of regression trees, in which each new tree is fitted to the residuals of the current ensemble, yielding a nonlinear mapping from input features to the target quantity. Using historical intervals from the archive underlying $\mathcal{S}$, a regressor is trained that maps day-ahead load and renewable generation forecasts, together with calendar features (hour of day and month), to the corresponding KE value~\eqref{eq:ke_fcr_forecast}. 
The unit-commitment pattern of the training year enters the forecast only through load and renewable generation levels, while calendar effects are captured by the hour- and month-level features; no explicit day-type indicator is used.
\begin{equation}
    \widehat{KE}_{t}
    =\Phi^{\mathrm{KE}}\bigl(
        \widehat{P}_{t}^{\mathrm{load}},\,
        \widehat{P}_{t}^{\mathrm{PV}},\,
        \widehat{P}_{t}^{\mathrm{wind}},\,
        h_{t},\,
        m_{t}
    \bigr)
\label{eq:ke_fcr_forecast}
\end{equation}

At the day-ahead stage, the FFR requirement is determined using the predicted kinetic energy $\widehat{KE}_{t}$ and the contingency magnitude $P^{\mathrm{cont}}$, typically corresponding to the largest infeed. The former is obtained from~\eqref{eq:ke_fcr_forecast}\footnote{Any alternative TSO or published forecasting method can replace $\Phi^{\mathrm{KE}}$ without changing the FFR request pipeline.}, while the latter depends on the availability of generation schedules before the FFR market gate closure. Accordingly, two practical cases are considered. 
If day-ahead energy market results and unit schedules are not yet available, $\widehat{KE}_{t}$ is obtained from the forecast, while $P^{\mathrm{cont}}$ is determined through a TSO risk assessment, analogous to FCR dimensioning (e.g., assuming a conservative fixed largest infeed)~\cite{ENTSOE_CE_SAFA_2024}. Conversely, if market participants are required to submit generation schedules reflecting the day-ahead market outcome before FFR clearing, as in Cyprus, $P^{\mathrm{cont}}$ can be directly determined from these schedules. In this case, kinetic energy can still be estimated using~\eqref{eq:ke_fcr_forecast}, particularly when subsequent market stages may alter unit commitments, making schedule-based inertia estimates less reliable than a dedicated forecast. Given the offline envelope and the chosen day-ahead inputs, the TSO dynamic FFR request for interval~$t$ is the surface lookup \eqref{eq:ffr_dynamic_request}. This closes the loop between the offline DSA-based sizing of~\cite{PANAGI2025101699} and the online market-dynamic capacity request from the TSO.
\begin{equation}
    R_{t}^{\mathrm{FFR}}
    = \widehat{P}^{\mathrm{FFR}}\!\left(\widehat{KE}_{t},\,\widehat{P}^{cont}_{t}\right)
    \label{eq:ffr_dynamic_request}
\end{equation}

\section{Risk-Aware Quantification of FFR Headroom}
\label{sec:method}

This section develops the aggregator-side methodology for the risk-aware quantification of the available FFR capacity headroom from energy storage and thermostatically flexible resources, subject to TSO certification requirements.

\subsection{Deterministic Resource Headroom Calculation}
\label{sec:deterministic_headroom}

As a first step, the available FFR headroom for each resource is quantified deterministically, assuming its scheduled operating power is known. The schedule denotes the DER baseline operating point and is not necessarily controllable by the aggregator. In practice, it reflects the users' normal behavior or the outcome of local energy management strategies. FFR reserve considers only the flexibility direction that supports the system during a frequency drop, namely a reduction of net consumption (for EVs, this may include a transition to V2G injection). Load-increasing flexibility is outside the scope of this paper, as no corresponding FFR market for over-frequency currently exists. Furthermore, as discussed in Section~\ref{sec:background}, FFR activations are rare and deliver only a negligible amount of energy. Consequently, state-of-energy and comfort limits, including battery SoC for storage-based resources, and domestic hot water tank or indoor temperatures for thermostatically controlled loads, are neglected in the FFR headroom quantification, and only power limits are retained. The proposed methodology remains applicable to more energy-intensive products (e.g., FCR or FRR) by reinstating these constraints as in~\cite{panagioptimal,panagi2026ifac}.

\subsubsection{Electric Vehicles}
\label{sec:ev_ffr_headroom}

Let $P_{ev,t}^{\mathrm{sch}}$ denote the scheduled charging power of $ev$ at time $t$ and $\overline{P}_{ev}^{\mathrm{dis}}$ its maximum V2G discharge power. Let $u_{ev,t}\in\{0,1\}$ indicate whether the EV is connected to the charger, and $v_{ev}\in\{0,1\}$ whether the vehicle and its charger are V2G capable.
When the EV is disconnected, no FFR can be delivered, $P_{ev,t}^{\mathrm{FFR}} = 0$ ($u_{ev,t}=0$). When connected, a V2G-capable EV can reduce its net power from the scheduled operating point $P_{ev,t}^{\mathrm{sch}}$ down to $-\overline{P}_{ev}^{\mathrm{dis}}$ (full V2G), whereas a unidirectional EV can only interrupt its charging, so that its headroom equals the scheduled charging power and depends on the charging policy that sets $P_{ev,t}^{\mathrm{sch}}$. The corresponding FFR headroom is therefore \eqref{eq:ev_ffr_headroom}.
\begin{equation}
    \label{eq:ev_ffr_headroom}
    P_{ev,t}^{\mathrm{FFR}} = u_{ev,t}\, (P_{ev,t}^{\mathrm{sch}} + v_{ev}\,\overline{P}_{ev}^{\mathrm{dis}})
\end{equation}

The same power-based headroom structure applies directly to stationary battery energy storage systems (BESS). The principal difference is that stationary storage has no connectivity constraint ($u\equiv 1$), so the availability binary disappears and the quantification reduces to the converter power limits alone. Consequently, FFR capability assessment for BESS is conceptually simpler than EVs.

\subsubsection{Heat Pumps}
\label{sec:hp_ffr_headroom}
For thermostatically controlled loads, FFR headroom is obtained solely by reducing electrical consumption relative to the scheduled operating point. Thus, if $P_{hp,t}^{\mathrm{sch}}$ denotes the scheduled HP electrical power, the FFR headroom equals the scheduled consumption,
\begin{equation}
    P_{hp,t}^{\mathrm{FFR}} = P_{hp,t}^{\mathrm{sch}}.
    \label{eq:hp_ffr_headroom}
\end{equation}

\subsection{Aggregator Model: Probabilistic Headroom Quantification and Risk-Aware Bidding}
\label{sec:probabilistic_headroom}

In practice, neither connectivity nor scheduled operating powers are known with certainty at the bidding stage. The deterministic resource maps of Section~\ref{sec:deterministic_headroom} must therefore be embedded in a probabilistic framework that yields, for each market period $t$, a reliable capacity offer from the EV and HP fleets. Several probabilistic methodologies can serve this purpose, including scenario sampling, chance-constrained optimization, and conditional quantile estimation, and the aggregator remains free to propose the procedure that best fits its portfolio and data availability. The TSO is responsible for reviewing and approving the proposed quantification method before the resulting bids are accepted in the FFR market, in line with the principle that reserve providers must document how offered capacity is derived and can be delivered. This pre-qualification process is similar to that used by the Danish TSO for the FCR-D market~\cite{10960637} (the P90 rule). 

\subsubsection{Historical Headroom Trajectories and Quantile Regression}
\label{sec:qr_ffr_bidding}

To translate the deterministic resource models into reliable day-ahead FFR bids, the \textit{Historical Quantile Regression Headroom (HQRH)} method is proposed. As illustrated in Algorithm~\ref{alg:risk_aware_ffr_bid} the method consists of three stages: (i)~map historical baselines to resource-level FFR headroom, (ii)~train technology-specific quantile models, and (iii)~sum the day-ahead technology quantiles to form the portfolio bid.

\begin{algorithm}[!ht]
\caption{Risk-aware day-ahead FFR bidding for an aggregator portfolio (HQRH)}
\label{alg:risk_aware_ffr_bid}
\begin{algorithmic}[1]
\REQUIRE Portfolio $\mathcal{A}=\mathcal{A}^{\mathrm{EV}}\cup\mathcal{A}^{\mathrm{HP}}$; historical/synthetic baselines $\{P_{i,\tau}^{\mathrm{sch}},u_{i,\tau}\}$; reliability $1-\varepsilon$; weather/calendar archives
\ENSURE Risk-aware offers $\{c_t\}$ for all day-ahead periods $t$

\STATE\textbf{Offline --- historical FFR headroom trajectories}
\FOR{each historical interval $\tau$}
    \FOR{each EV $ev\in\mathcal{A}^{\mathrm{EV}}$}
        \STATE $P_{ev,\tau}^{\mathrm{FFR}} \leftarrow u_{ev,\tau}\bigl(P_{ev,\tau}^{\mathrm{sch}}+v_{ev}\,\overline{P}_{ev}^{\mathrm{dis}}\bigr)$
        \COMMENT{cf.~\eqref{eq:ev_ffr_headroom}}
    \ENDFOR
    \FOR{each HP $hp\in\mathcal{A}^{\mathrm{HP}}$}
        \STATE $P_{hp,\tau}^{\mathrm{FFR}} \leftarrow P_{hp,\tau}^{\mathrm{sch}}$
        \COMMENT{cf.~\eqref{eq:hp_ffr_headroom}}
    \ENDFOR
    \STATE $P_{\mathcal{A}^{\mathrm{EV}},\tau}^{\mathrm{FFR}} \leftarrow \! \! \sum_{ev} P_{ev,\tau}^{\mathrm{FFR}}$,\;
          $P_{\mathcal{A}^{\mathrm{HP}},\tau}^{\mathrm{FFR}} \leftarrow \! \! \sum_{hp} P_{hp,\tau}^{\mathrm{FFR}}$
        \COMMENT{cf.~\eqref{eq:tech_ffr_sum}}
    \STATE Build features $x_{\tau}^{\mathrm{EV}}$, $x_{\tau}^{\mathrm{HP}}$ via~\eqref{eq:ffr_features}
\ENDFOR

\STATE\textbf{Offline --- train QRF quantile models~\cite{meinshausen2006quantile}}
\STATE Form $\mathcal{D}^{\mathrm{EV}}$ and $\mathcal{D}^{\mathrm{HP}}$ as in~\eqref{eq:ffr_train_sets}
\STATE Train QRF$^{\mathrm{EV}}$ on $\mathcal{D}^{\mathrm{EV}}$ and QRF$^{\mathrm{HP}}$ on $\mathcal{D}^{\mathrm{HP}}$
\STATE \COMMENT{Each forest stores leaf-wise empirical distributions of FFR headroom}

\STATE\textbf{Day-ahead --- apply models and bid}
\FOR{each market period $t$ of the next day}
    \STATE Forecast / assemble $x_{t}^{\mathrm{EV}}$ and $x_{t}^{\mathrm{HP}}$ \eqref{eq:ffr_features}
    \STATE $c_{t}^{\mathrm{EV}} \leftarrow$ $\varepsilon$-quantile from QRF$^{\mathrm{EV}}(x_{t}^{\mathrm{EV}})$
    \STATE $c_{t}^{\mathrm{HP}} \leftarrow$ $\varepsilon$-quantile from QRF$^{\mathrm{HP}}(x_{t}^{\mathrm{HP}})$
    \STATE $c_t \leftarrow c_{t}^{\mathrm{EV}}+c_{t}^{\mathrm{HP}}$
        \COMMENT{cf.~\eqref{eq:ffr_quantile_bid}}
\ENDFOR
\RETURN $\{c_t\}$
\end{algorithmic}
\end{algorithm}

\paragraph{Step 1 --- Historical FFR trajectories}
Whenever operational measurements (or synthetics) are available, the deterministic maps~\eqref{eq:ev_ffr_headroom} and~\eqref{eq:hp_ffr_headroom} are applied retrospectively to each past interval~$\tau$. Resource-level series are then summed within each technology class \eqref{eq:tech_ffr_sum}, to obtain the EV and HP portfolio targets used for quantile learning.
\begin{equation}
    P_{\mathcal{A}^{\mathrm{EV}},\tau}^{\mathrm{FFR}}=\sum_{ev\in\mathcal{A}^{\mathrm{EV}}} P_{ev,\tau}^{\mathrm{FFR}}, \qquad
    P_{\mathcal{A}^{\mathrm{HP}},\tau}^{\mathrm{FFR}} = \sum_{hp\in\mathcal{A}^{\mathrm{HP}}} P_{hp,\tau}^{\mathrm{FFR}},
    \label{eq:tech_ffr_sum}
\end{equation}

\paragraph{Step 2 --- Feature sets and model choice}
Because EV and HP uncertainty are driven by different exogenous factors, two separate feature vectors are used \eqref{eq:ffr_features}.
\begin{equation}
    \begin{aligned}
        x_{\tau}^{\mathrm{EV}}
        &=\bigl(h_{\tau},\,d_{\tau}\bigr), \\
        x_{\tau}^{\mathrm{HP}}
        &=\bigl(T_{\tau}^{\mathrm{out}},\, s_{\tau},\, h_{\tau},\, d_{\tau},\, P_{\mathcal{A}^{\mathrm{HP}},\tau-48}^{\mathrm{FFR}}\bigr),
    \end{aligned}
    \label{eq:ffr_features}
\end{equation}
where $h_{\tau}$ is the hour of day and $d_{\tau}$ the day type (weekday/weekend). For EVs, these calendar features alone capture the dominant commuting-and-home-availability cycle. For HPs, $T_{\tau}^{\mathrm{out}}$ is the day-ahead weather driver; $s_{\tau}$ is a season indicator; and $P_{\mathcal{A}^{\mathrm{HP}},\tau-48}^{\mathrm{FFR}}$ is the realised headroom 48~h earlier, which is already known at day-ahead gate closure. All of these inputs align with the main drivers of mobility and building-load uncertainty.

\begin{figure}
\centering
\includegraphics[width=\columnwidth]{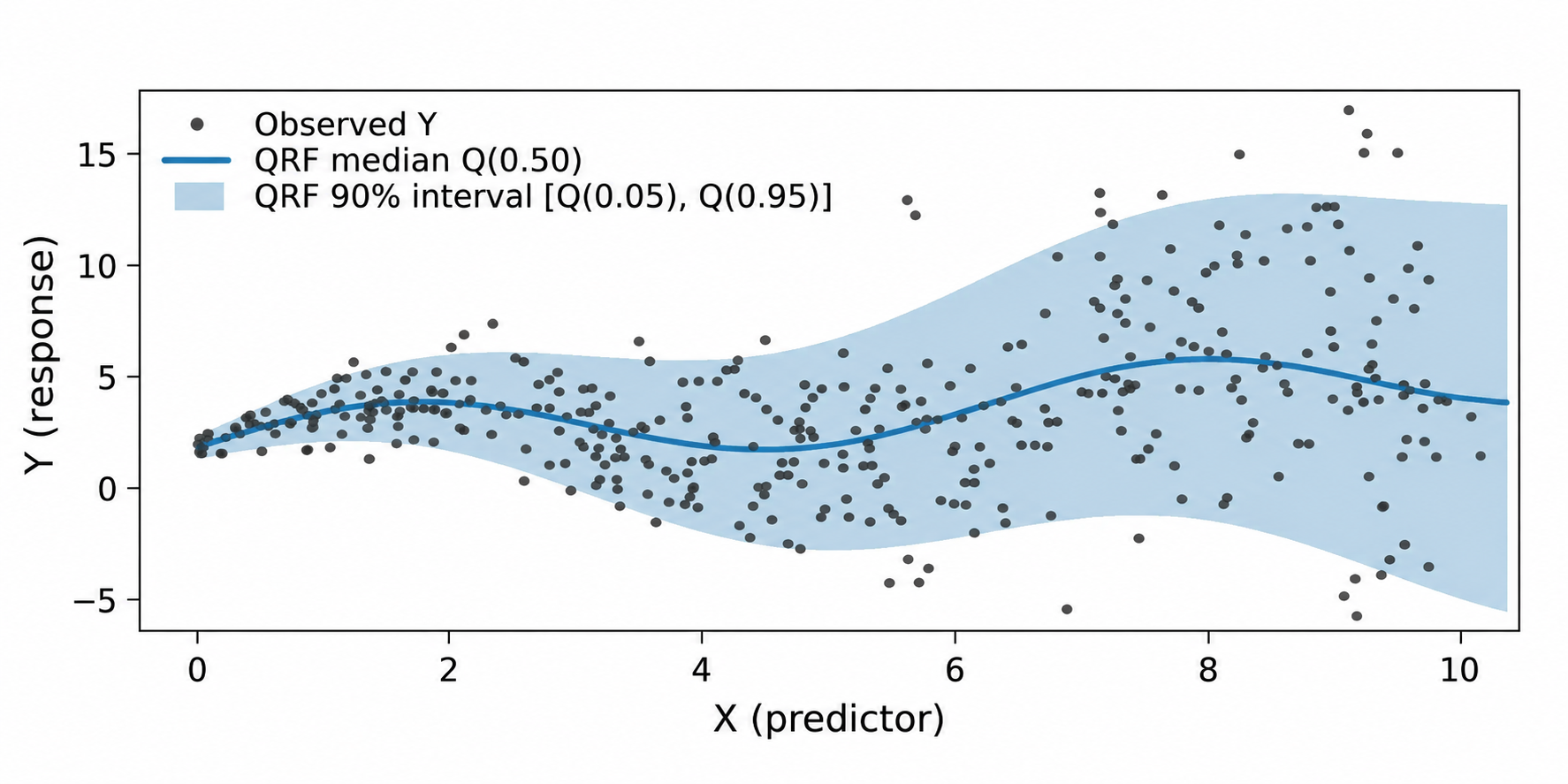}
\caption{Illustrative QRF prediction: median $Q_{0.50}$ and $90\%$ band $[Q_{0.05},Q_{0.95}]$. Interval width varies with $x$, capturing uncertainty \cite{meinshausen2006quantile}.}
\label{fig:qrf_heteroscedastic}
\end{figure}

Among the quantile-regression families surveyed in~\cite{xu2023quantile}, including GAQR, QRNN, SVQR, QRF, and deep QR models, QRF~\cite{meinshausen2006quantile} is selected. A conventional random forest estimates only the conditional mean of a response $Y$ given features $x$. QRF keeps the same forest structure but stores, at each leaf, the empirical collection of training responses rather than their average. For a query $x$, leaf assignments across $B$ trees induce weights $w_i(x)$ on the training samples $\{Y_i\}_{i=1}^{n}$, and the conditional distribution is recovered as the weighted empirical CDF~\cite{meinshausen2006quantile} \eqref{eq:qrf_cdf}. Any quantile is therefore available from a single trained forest. Fig.~\ref{fig:qrf_heteroscedastic} illustrates the practical consequence: unlike a mean-only regression, QRF yields a prediction band whose width adapts to local uncertainty. In the present setting, $Y$ is the technology-level FFR headroom of~\eqref{eq:tech_ffr_sum} and $x$ is the corresponding feature vector in~\eqref{eq:ffr_features}; Step~3 then uses a lower quantile of this distribution as the risk-aware bid~\eqref{eq:qrf_cdf_specific_quantile}.
\begin{align}
    \hat{F}(y\mid x) = \sum_{i=1}^{n} w_i(x)\,\mathbf{1}\{Y_i\le y\},
    \label{eq:qrf_cdf} \\
    \hat{Q}_{\alpha}(x) = \inf\bigl\{y:\hat{F}(y\mid x)\ge\alpha\bigr\}.
    \label{eq:qrf_cdf_specific_quantile}
\end{align}
The weight $w_i(x)=\frac{1}{B}\sum_{b=1}^{B}\mathbf{1}\{X_i\in\ell_b(x)\}/|\ell_b(x)|$ is the average over the $B$ trees of the indicator that training sample $i$ falls in the leaf $\ell_b(x)$ that contains $x$, normalised by the number of training samples in that leaf, so that $\sum_{i} w_i(x)=1$. QRF is adopted because it (i)~handles nonlinear tabular drivers without strong distributional assumptions, (ii)~is computationally moderate for aggregator-scale datasets, (iii)~avoids the heavy tuning of QRNN/SVQR/deep QR, and (iv)~returns the lower quantile needed for risk-aware bidding.

\paragraph{Step 3 --- Training and day-ahead application}
Let $\varepsilon\in(0,1)$ be the admissible shortfall probability. Two QRF models are trained offline on the technology datasets~\eqref{eq:ffr_train_sets}, with responses equal to the fleet headrooms of~\eqref{eq:tech_ffr_sum}. Day-ahead, each model evaluates the conditional lower quantile~\eqref{eq:qrf_cdf_specific_quantile} at the forecast features, which yields the technology offers~\eqref{eq:ffr_quantile_tech}.
\begin{equation}
    \mathcal{D}^{\mathrm{EV}}
    =    \bigl(x_{\tau}^{\mathrm{EV}},\,P_{\mathcal{A}^{\mathrm{EV}},\tau}^{\mathrm{FFR}}\bigr), \quad \mathcal{D}^{\mathrm{HP}} =\bigl(x_{\tau}^{\mathrm{HP}},\,P_{\mathcal{A}^{\mathrm{HP}},\tau}^{\mathrm{FFR}}\bigr),
    \label{eq:ffr_train_sets}
\end{equation}
\begin{equation}
    \begin{aligned}
        c_{t}^{\mathrm{EV}}&=
        \hat{Q}_{\varepsilon}\!\bigl(x_{t}^{\mathrm{EV}}\bigr)
        =
        Q_{\varepsilon}\!\bigl(P_{\mathcal{A}^{\mathrm{EV}},t}^{\mathrm{FFR}}\,\big|\,x_{t}^{\mathrm{EV}}\bigr),
        \\
        c_{t}^{\mathrm{HP}}
        &=
        \hat{Q}_{\varepsilon}\!\bigl(x_{t}^{\mathrm{HP}}\bigr)
        =
        Q_{\varepsilon}\!\bigl(P_{\mathcal{A}^{\mathrm{HP}},t}^{\mathrm{FFR}}\,\big|\,x_{t}^{\mathrm{HP}}\bigr).
    \end{aligned}
    \label{eq:ffr_quantile_tech}
\end{equation}
Here $x_{\tau}^{\mathrm{(\cdot)}}$ are historical features used for training, whereas $x_{t}^{\mathrm{(\cdot)}}$ are the corresponding day-ahead (forecast) features used at bidding time. Choosing $c=\hat{Q}_{\varepsilon}$ enforces the technology-wise chance constraints~\eqref{eq:ffr_chance_constraint} under the learned conditional laws. 

\begin{equation}
    \begin{aligned}
        \mathbb{P}\!\bigl(P_{\mathcal{A}^{\mathrm{EV}},t}^{\mathrm{FFR}}\ge c_{t}^{\mathrm{EV}}\bigm|x_{t}^{\mathrm{EV}}\bigr)
        &\ge 1-\varepsilon,
        \\
        \mathbb{P}\!\bigl(P_{\mathcal{A}^{\mathrm{HP}},t}^{\mathrm{FFR}}\ge c_{t}^{\mathrm{HP}}\bigm|x_{t}^{\mathrm{HP}}\bigr)
        &\ge 1-\varepsilon.
    \end{aligned}
    \label{eq:ffr_chance_constraint}
\end{equation}
The \emph{portfolio} offer is obtained after the two quantiles are evaluated, by summing the technology bids~\eqref{eq:ffr_quantile_bid}.
\begin{equation}
    c_{t}
    =
    c_{t}^{\mathrm{EV}}+c_{t}^{\mathrm{HP}},
    \label{eq:ffr_quantile_bid}
\end{equation}

\subsection{Complete Proposed Methodology}
\label{sec:complete_methodology}

\begin{figure}[!ht]
\centering
\includegraphics[width=0.9\columnwidth]{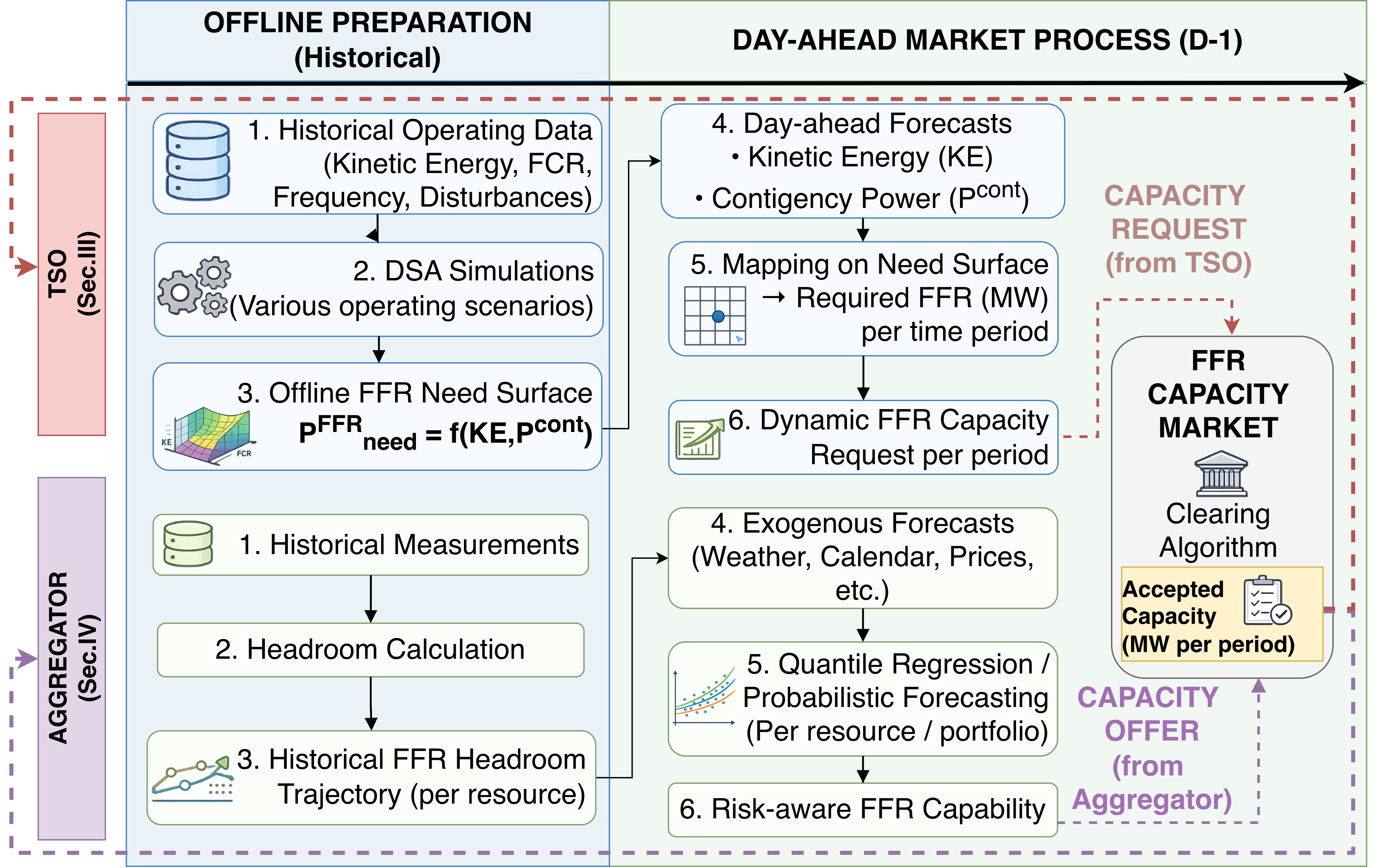}
\caption{End-to-end proposed methodology. Upper row: TSO offline need-surface construction and day-ahead dynamic capacity request. Lower row: aggregator historical headroom trajectories and risk-aware day-ahead offer. Both feed the FFR capacity market clearing on the day-ahead horizon.}
\label{fig:complete_methodology}
\end{figure}

Fig.~\ref{fig:complete_methodology} summarizes the end-to-end structure proposed in this paper. The upper branch is the TSO request pipeline of Section~\ref{sec:dynamic_requests}; the lower branch is the aggregator offer pipeline of Sections~\ref{sec:deterministic_headroom}--\ref{sec:probabilistic_headroom}. The two branches meet at the FFR capacity market, where requested and offered volumes are cleared for each interval~$t$. Offline, the TSO builds a secure need surface $\widehat{P}^{\mathrm{FFR}}(KE,P^{\mathrm{cont}})$. On the day-ahead horizon, system forecasts are mapped to $\widehat{KE}_t$, and the dynamic capacity request follows by surface lookup~\eqref{eq:ffr_dynamic_request}.
In parallel, the aggregator maps customer baselines to resource FFR via~\eqref{eq:ev_ffr_headroom}--\eqref{eq:hp_ffr_headroom}, forms technology targets~\eqref{eq:tech_ffr_sum}, trains two QRF models on~\eqref{eq:ffr_features}, and bids the summed quantiles~\eqref{eq:ffr_quantile_bid} targeting~\eqref{eq:ffr_chance_constraint}.

The TSO requests $R_t^{\mathrm{FFR}}$ and the capacity--price offers of all eligible providers $a\in\mathcal{A}^{\mathrm{FFR}}$ then enter a day-ahead FFR capacity clearing problem that minimizes the TSO procurement cost~\eqref{eq:ffr_cost_clearing}. Let $c_{a,t}$ and $\pi_{a,t}$ denote, respectively, the offered capacity and bid price of provider~$a$ in interval~$t$, $\Delta t$ be the market-interval duration, and $p_{a,t}$ be the accepted capacity. The equality constraint enforces coverage of the dynamic request $R_t^{\mathrm{FFR}}$, and the offer bounds ensure that cleared volumes do not exceed the risk-aware capabilities. Because the objective is separable in $t$ and settlement is pay-as-bid, \eqref{eq:ffr_cost_clearing} decomposes into a per-interval merit order in which offers are accepted in increasing price order until the request is met.

\begin{equation}
    \begin{aligned}
        \min_{\{p_{a,t}\}} &\quad\sum_{t} \sum_{a\in\mathcal{A}^{\mathrm{FFR}}} \pi_{a,t}\, p_{a,t}\Delta t \\
        \text{s.t.}
        &\quad \sum_{a\in\mathcal{A}^{\mathrm{FFR}}} p_{a,t} = R_{t}^{\mathrm{FFR}},
        && \forall t\in\mathcal{T}, \\
        &\quad 0 \le p_{a,t} \le c_{a,t}, && \forall a,t.
    \end{aligned}
    \label{eq:ffr_cost_clearing}
\end{equation}

\section{Case Study}
\label{sec:case_study}

All case-study data used in this paper, including synthetic EV and HP load profiles, day-ahead FFR capacity-price offers, and the 2025 TSO FFR capacity requests, are publicly available~\cite{panagi_2026_23556477}, with the exception of the Cyprus power-system operating points and RMS model data, which are confidential.

\subsection{Power System and Dynamic Model}
\label{sec:cyprus_system}
\begin{figure}
    \centering
    \includegraphics[width=0.8\columnwidth]{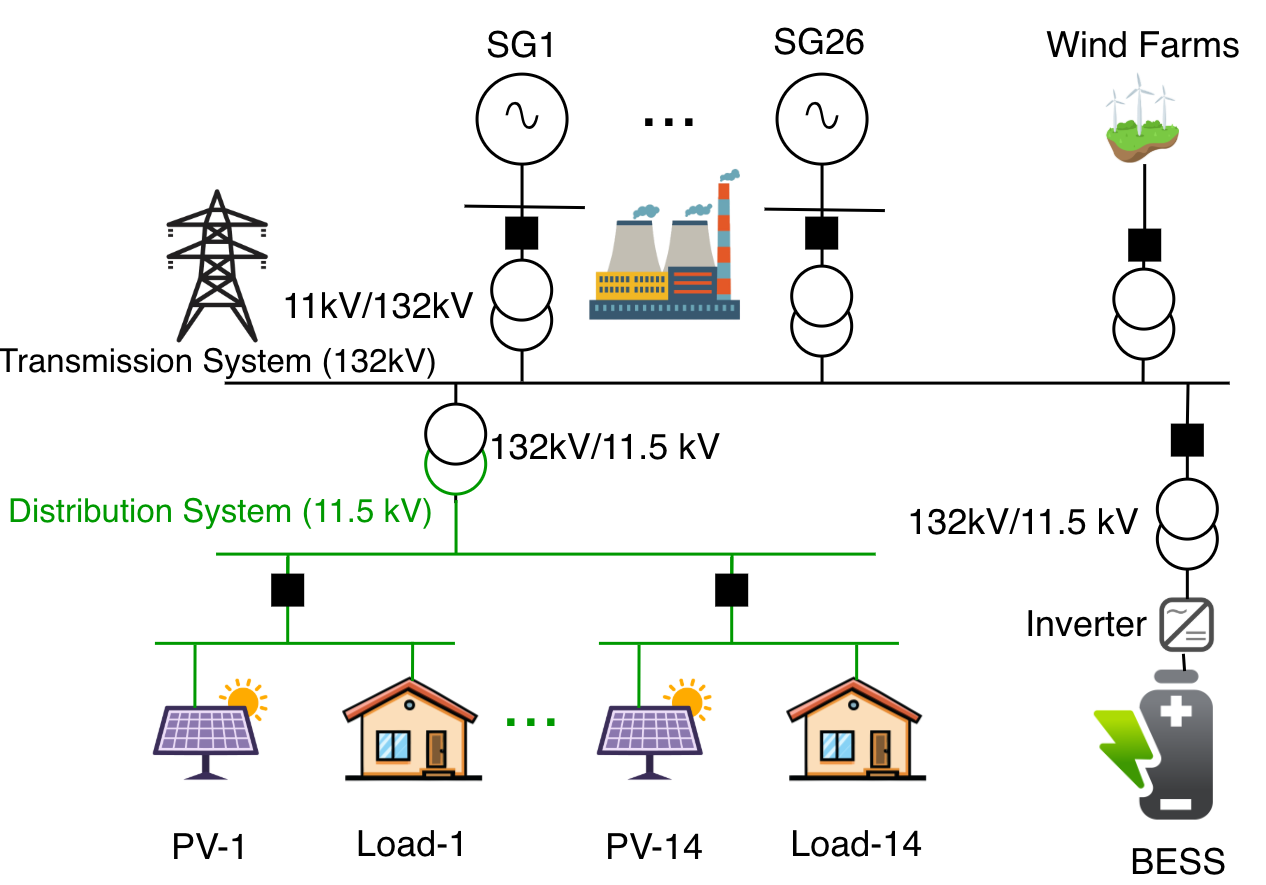}
    \caption{One-line diagram of the Cyprus power system.}
    \label{fig:cyprus_system_one_line}
\end{figure}

The case study is based on the realistic, low-inertia, islanded Cyprus power system, as depicted in Fig.~\ref{fig:cyprus_system_one_line}. The model used here is the same operational dynamic representation employed by the Cyprus TSO for frequency-related studies. A reduced network topology with detailed dynamic models of synchronous generation, loads, and renewable resources, is implemented in DIgSILENT PowerFactory~\cite{argyrou2024synergy,PANAGI2025101699}. The system comprises 26 conventional SG-based units (steam, gas, and diesel), all equipped with AVRs, governors, and PSS models consistent with TSO data, as well as wind farms and 14 groups of aggregated distributed PV and loads. 

An aggregated BESS with FFR capability is also implemented to represent the aggregated FFR response. Moreover, the Cyprus power system currently relies on UFLS to limit excessive FCR commitments, which would otherwise require increased operation of conventional generators and, consequently, higher RES curtailment, as FCR is predominantly provided by conventional generation. Therefore, an additional aggregated BESS with FCR-only capability is introduced across the historical operating points to cover any FCR shortfall and enable the assessment of FFR requirements under adequate FCR availability. 
The first UFLS stage is at $49\,\mathrm{Hz}$, which is also the nadir target used for FFR sizing. Further details of the power system model are documented in~\cite{argyrou2024synergy,PANAGI2025101699}. 

\begin{figure}
    \centering
    \includegraphics[width=0.8\linewidth]{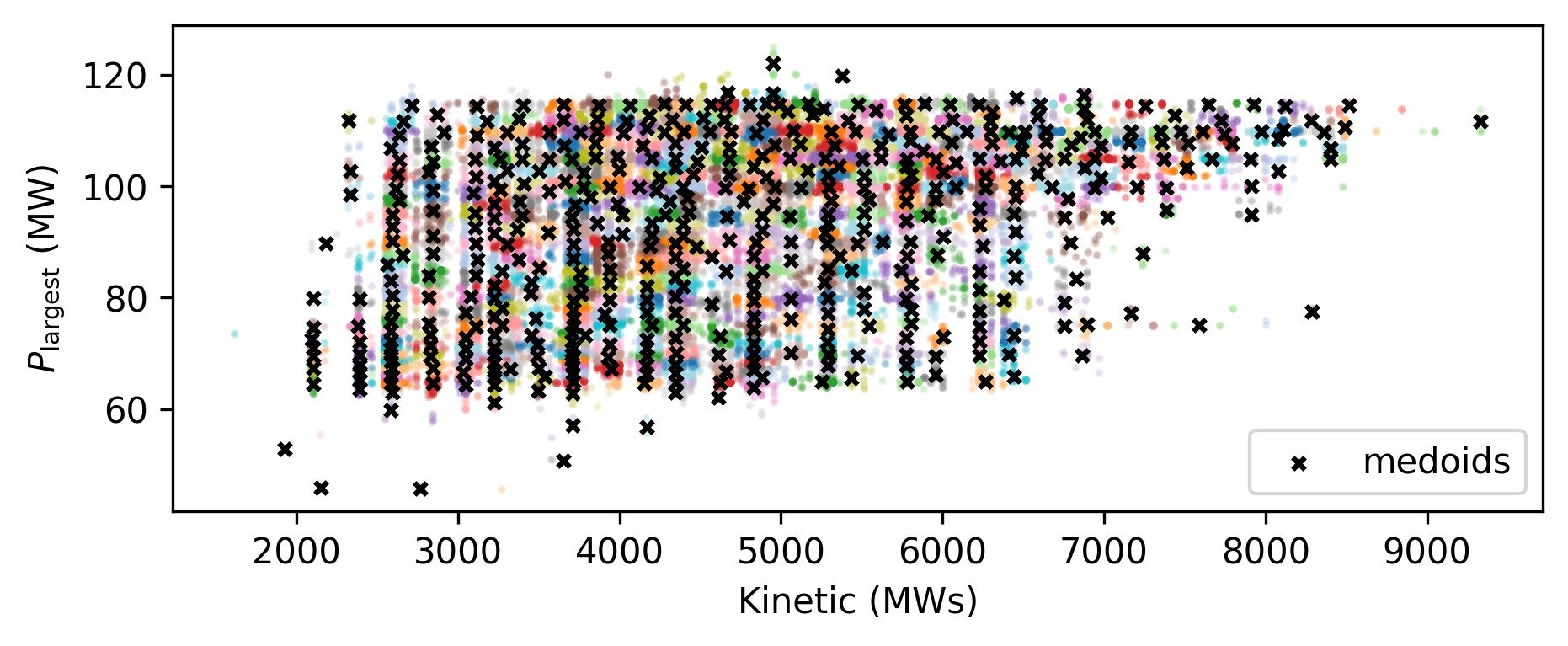}
    \caption{Representative 2024 operating points selected by k-means clustering with medoid selection.}
    \label{fig:clustering_scenarios}
\end{figure}

The day-ahead TSO request~$R_t^{\mathrm{FFR}}$ is determined as described in Section~\ref{sec:dynamic_requests}. The offline archive~$\mathcal{S}$ and the FFR need surface are constructed using historical operating data from 2024, while data from 2025 are used for validation. The kinetic-energy forecast is an XGBoost regressor with 400 trees, depth 4, and learning rate 0.05. With a 15-minute resolution, the 2024 dataset contains 35,136 operating points. Performing RMS simulations and quantifying the FFR requirement for every operating point would be computationally intensive. Therefore, 500 representative operating points are selected using k-means clustering with medoid selection, considering KE and the largest contingency magnitude as the main features, following an approach similar to the publicly available implementation in~\cite{SPSL_TSOC_Data_Analysis}. The resulting representative operating points are shown in Fig.~\ref{fig:clustering_scenarios}. Over these points, $\Delta_{\mathrm{KE}}=(KE_{\max}-KE_{\min})/14=529\,\mathrm{MWs}$ and $\Delta_{\mathrm{cont}}=(P^{\mathrm{cont}}_{\max}-P^{\mathrm{cont}}_{\min})/10=7.65\,\mathrm{MW}$.

\subsection{Aggregator Portfolio and Synthetic Baselines}
\label{sec:aggregator_portfolio}

The HP baselines consist of synthetic electrical demand profiles derived from physics-based building simulations. A two-storey, $201.6~\mathrm{m}^2$ detached reference dwelling, representative in size of Cypriot single-family housing~\cite{cystat_housing}, is modeled in EnergyPlus 26.1~\cite{crawley2001energyplus}. Three construction vintages are considered: 1980, 2003, and recent. Their envelopes, internal gains, thermostat schedules and airtightness follow the official reference dwellings of the 2023 Cyprus cost-optimal study~\cite{cyprus_costoptimal2023} and the NCM residential activity database~\cite{ncm_activity_db}. For each dwelling, the thermal demand of the five conditioned rooms is simulated using the EnergyPlus ideal loads approach. The resulting thermal demand is converted into the electrical demand of one equivalent reversible air-to-air heat pump per dwelling using the ambient-temperature-dependent performance factors $COP(T)=3.0e^{0.016T}$ and $EER(T)=11.0e^{-0.029T}$~\cite{olympios2025stochastic}. The modeling framework follows~\cite{kourougianni2024strategies}, extended with a seeded stochastic model in which a persistent timing preference ($\pm60$~min), a day-to-day jitter ($\pm30$~min) and thermostat offsets ($\pm1$~K) perturb the schedules of each dwelling. In total, $N_{\mathrm{HP}}=1000$ distinct households are simulated, distributed as 334/333/333 across the three construction vintages. The same fleet is simulated for the actual meteorological years 2024 (QRF training) and 2025 (out-of-sample test), with weather from the ERA5 reanalysis via Open-Meteo~\cite{hersbach2020era5,zippenfenig2023openmeteo}.

EV baselines are more subtle. A common approach in reserve bidding literature is to treat historical charger plug-in/out intervals as availability~\cite{10960637}. For FFR (and any product that remunerates for being connected), this is problematic. Such traces reflect past charging needs and habits, not the mobility windows in which the vehicle can remain connected. Revenue-maximizing owners have an incentive to stay plugged in beyond energy replenishment, so equating historical connections with future availability systematically underestimates the FFR headroom.

To address this drawback, \emph{mobility} is decoupled from \emph{charging} and a home-charging fleet of $N^{\mathrm{EV}}=1000$ vehicles is synthesized over a continuous year at $15\,\mathrm{min}$ resolution. Home departure/arrival times and tour distances are sampled from empirical home-based vehicle tours extracted from the Victorian Integrated Survey of Travel and Activity raw trip file, obtained from the open travel dataset package of~\cite{dimonti_2026_19364191}. For each EV-day, the sampled-away window defines the binary availability $u_{ev,t}=1$ when at home and $0$ otherwise. Round-trip energy is deducted from the battery at departure ($0.18\,\mathrm{kWh/km}$). While at home, vehicles charge at $\overline{P}^{\mathrm{ch}}=7.4\,\mathrm{kW}$ with a SoC hysteresis policy (start below $50\%$, stop at $90\%$); battery capacities are drawn uniformly in $[40,75]\,\mathrm{kWh}$. The resulting series $u_{ev,t}$ and $P_{ev,t}^{\mathrm{sch}}=P_{ev,t}^{\mathrm{ch}}$ feed the deterministic FFR map~\eqref{eq:ev_ffr_headroom}, with V2G discharge rating taken equal to the charger rating ($\overline{P}_{ev}^{\mathrm{dis}}=7.4\,\mathrm{kW}$) for the 20\% of the EVs that are V2G capable ($v_{ev}=1$). The HP forest uses 100 trees, depth 12 and a minimum leaf of 15 samples; the EV forest uses 200 trees, depth 12 and a minimum leaf of 5.

\subsection{Bidding Prices Assumptions}
\label{sec:ffr_price_assumptions}

Proxy bid prices are constructed from Fingrid~\cite{fingrid_ffr} hourly FFR clearances over 2024--2025. Only \emph{non-zero} clearance prices are retained ($n=2438$ active hours), corresponding to intervals in which the product was actually procured; these hours are concentrated in low-inertia periods, so the sample mean $\mu$ is conditional rather than unconditional. Let $\mu$ and $\sigma$ denote the sample mean and standard deviation of the non-zero Fingrid prices. For each market interval~$t$ of the aggregator study horizon, a price $\pi_t\sim\mathcal{N}(\mu,\sigma^2)$ is drawn i.i.d.\ (no autocorrelation); negative draws are redrawn and clipped at $0.01$~\euro/MW/h. Three scenarios share the same underlying draws and differ only by a scale factor $\pi_t^{(s)}=\alpha_s\,\pi_t$, $\alpha_s\in\{0.85,\ 1.00,\ 1.15\}$. The resulting prices are summarized in Table~\ref{tab:ffr_price_scenarios}, interpreted as capacity payments in \euro/MW/h, and used only for economic illustration of the risk-aware quantity offers. For the subsequent FFR capacity-market clearing, the 1000-unit HP and EV aggregator templates are replicated to represent multiple independent aggregators, ensuring that the total offered capacity exceeds the peak TSO requirement by 5\%. Aggregators within the same technology class follow identical FFR capacity profiles, while their bid prices vary across pricing scenarios. 

\begin{table}
\centering
\caption{Proxy FFR capacity-price scenarios (mean/std) based on Fingrid clearances for 2024--2025. }
\label{tab:ffr_price_scenarios}

\setlength{\tabcolsep}{3.5pt}
\begin{tabular}{@{}llcrr@{}}
\toprule
Scenario & Role & Scale $\alpha_s$ & Mean [€/MW/h] & Std [€/MW/h] \\
\midrule
$S_{-15\%}$ & Cheaper & $0.85$ & $24.7$ & $14.5$ \\
$S_{\mathrm{central}}$ & Baseline $\mathcal{N}(\mu,\sigma^2)$ & $1.00$ & $29.1$ & $17.1$ \\
$S_{+15\%}$ & More expensive & $1.15$ & $33.4$ & $19.6$ \\
\bottomrule
\end{tabular}
\end{table}

\section{Numerical Results}
\label{sec:Results_Discussions}

This section presents the results of the Cyprus case study along the three stages of the proposed framework, followed by sensitivity analyses. Specifically, the results address (i)~dynamic TSO FFR capacity requirements, (ii)~risk-aware FFR offers from EV and HP aggregators, (iii)~day-ahead FFR market clearing and its relation to a UFLS VoLL benchmark, and (iv)~sensitivity analyses to assess the robustness of the proposed framework.

\subsection{Dynamic FFR Capacity Requests}
\label{sec:results_tso_requests}

Table~\ref{tab:ke_fcr_plargest_metrics} summarizes the training and test performance of the gradient boosting model for kinetic energy forecasting in terms of MAE, RMSE, MAPE, and $R^{2}$. Overall, the model demonstrates good predictive accuracy, achieving an $R^{2}$ of 0.889 on the test set. 
The offline need surface, calculated using 2024 historical operating points, is shown in Fig.~\ref{fig:ffr_need_maximum}. The results reveal that FFR requirements generally increase with larger contingency magnitudes and lower kinetic energy levels. 

\begin{table}[b]
\centering
\caption{Forecasting performance of the XGBoost kinetic energy forecasting model.}
\label{tab:ke_fcr_plargest_metrics}
\begin{tabular}{lcccc}
\toprule
Target & MAE (MWs) & RMSE (MWs) & MAPE (\%) & $R^{2}$ \\
\midrule
Kinetic energy (train) & 260.9 & 339.0 & 6.35 & 0.938 \\
Kinetic energy (test)  & 304.2 & 389.6 & 7.11 & 0.889 \\
\bottomrule
\end{tabular}
\end{table}

\begin{figure}
\centering
\includegraphics[width=\columnwidth]{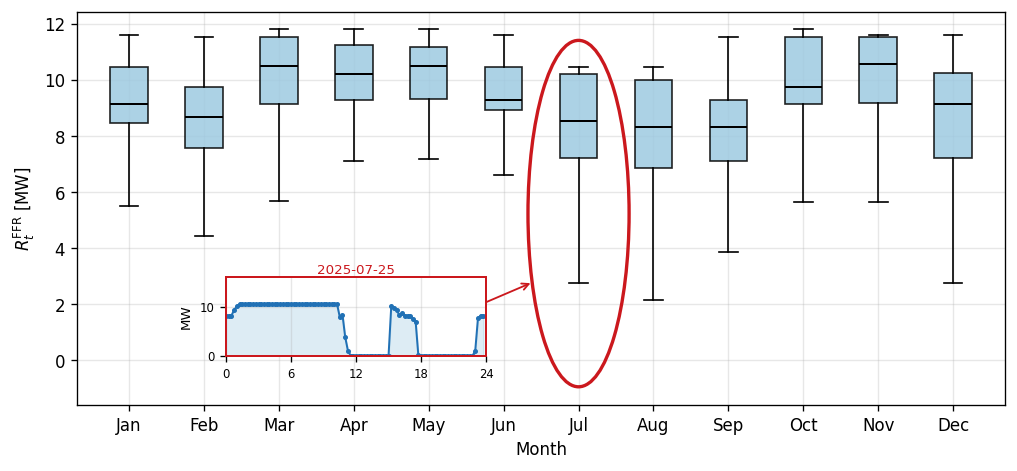}
\caption{Monthly distribution of 2025 dynamic FFR capacity requests. The inset shows the 15-min profile on $25$~July~2025.}
\label{fig:ffr_requirements}
\end{figure}

To determine the dynamic FFR capacity requirements, the proposed framework is applied to the 2025 operating conditions, which serve as an unseen test year. For each 15-minute interval, the forecasted kinetic energy is used as input to the need surface constructed from 2024 historical data. In addition, the contingency magnitude ($P^{\mathrm{cont}}$) is assumed to be known from the post-clearing market schedule, consistent with the Cyprus market framework. These inputs determine the FFR capacity requested by the TSO for each interval, as illustrated in Fig.~\ref{fig:ffr_requirements}. Over the year, the requirements average approximately $9.2\,\mathrm{MW}$, ranging from zero to $11.8\,\mathrm{MW}$. Summer months exhibit greater variability and more frequent intervals with low FFR requirements, whereas autumn and winter show a narrower distribution concentrated at higher values. 
This seasonal pattern follows from the commitment of synchronous units: the high summer demand keeps more conventional units online and raises KE, whereas the low-load autumn and winter days, with midday PV displacing further units, leave the system with lower KE and hence a higher FFR need. The inset highlights a representative summer day, demonstrating the pronounced intra-day variations that would not be captured by static annual or seasonal reserve procurement.

\subsection{Risk-Aware Aggregator Offers}
\label{sec:results_qrf}

\begin{figure}
    \centering
    \includegraphics[width=0.8\columnwidth]{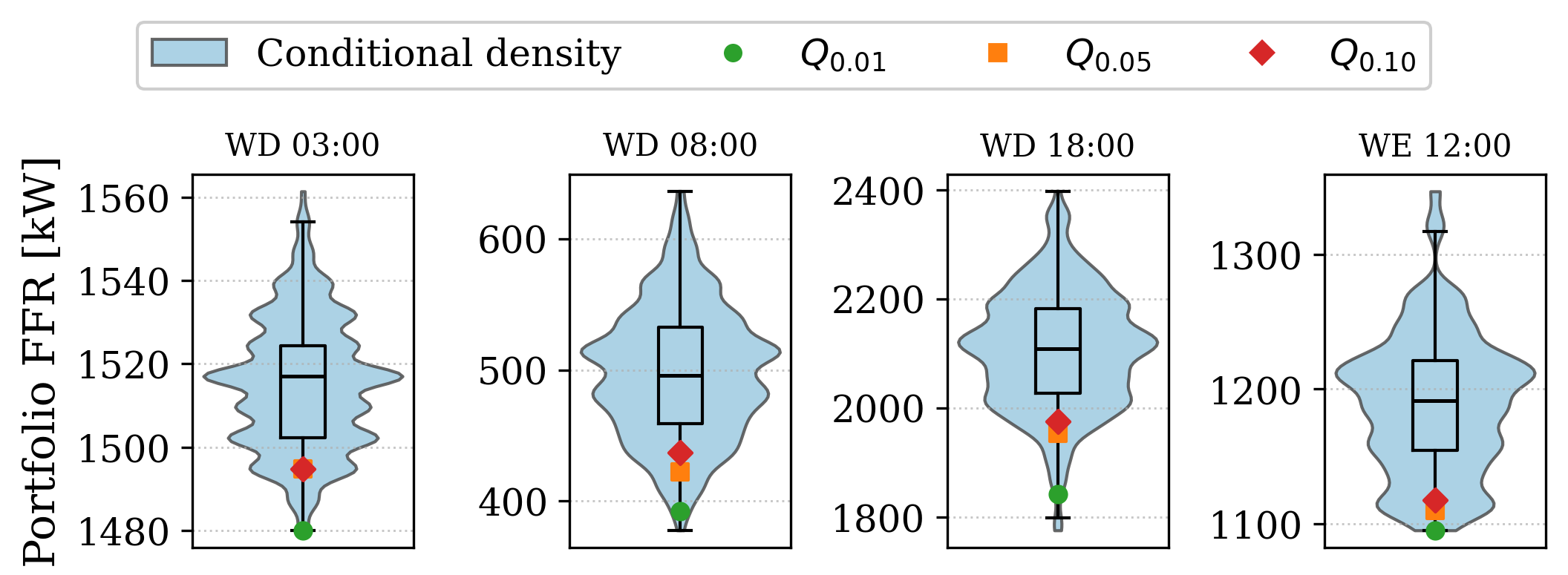}\\[0.4em]
    \includegraphics[width=0.8\columnwidth]{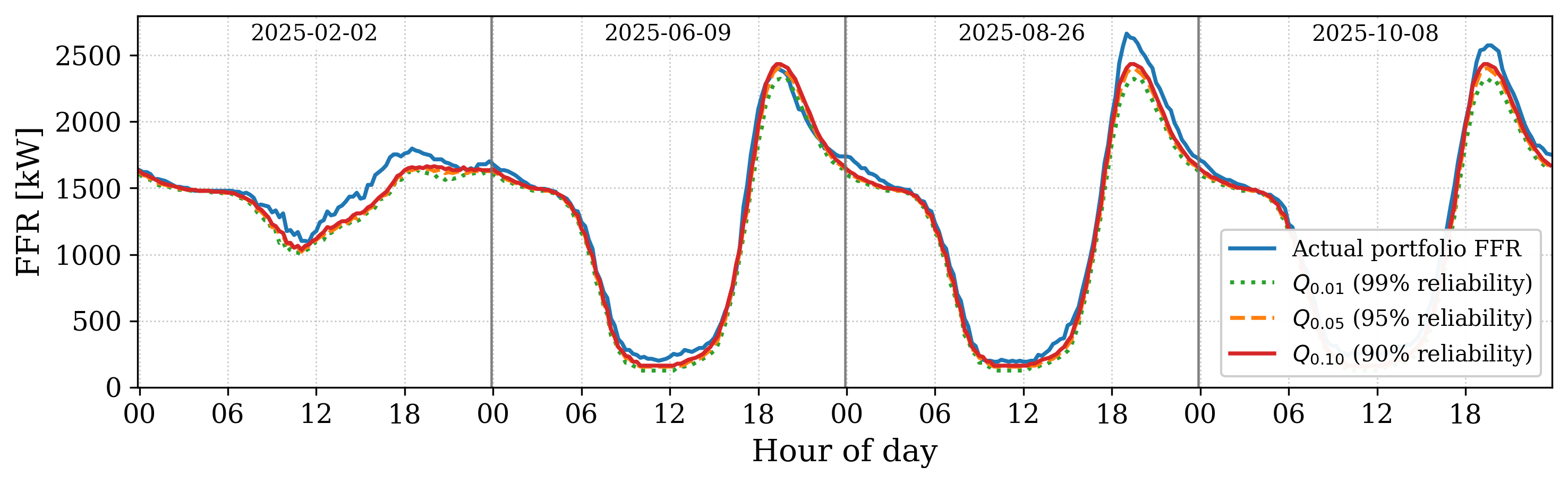}
    \caption{EV QRF results. Top: conditional distributions of portfolio FFR at selected hours of day, with lower-tail bids $Q_{0.01}$, $Q_{0.05}$, and $Q_{0.10}$ marked. Bottom: day-ahead quantile bids versus realized portfolio FFR on four test days.}
    \label{fig:results_qrf_ev}
\end{figure}

For each FFR provider technology, lower-tail bids $Q_{0.01}$, $Q_{0.05}$, and $Q_{0.10}$ (target reliabilities $99\%$, $95\%$, and $90\%$) are evaluated on the 15-min portfolio FFR series. Figure~\ref{fig:results_qrf_ev}~(a) shows EV conditional distributions at representative hours, with the three lower quantiles marked. Night-time headroom is high and tightly concentrated around the maximum V2G power ($\sim1.48\,\mathrm{MW}$, the 200 V2G-capable vehicles discharging at $7.4\,\mathrm{kW}$) with some small additional capacity due to late-night charging; the remaining 800 unidirectional EVs contribute only while they are charging. Morning commuting hours exhibit much lower and more dispersed FFR capability, consistent with reduced home availability. Evening hours are the most dispersed, reflecting the overlap of high connectivity and peak charging. Fig.~\ref{fig:results_qrf_ev}~(b) compares day-ahead bids against realized portfolio FFR on a random four test days. Offers track the daily cycle and remain below the actual almost everywhere, with only rare violations, as required for risk-aware capacity commitment.

\begin{figure}
    \centering
    \includegraphics[width=0.8\columnwidth]{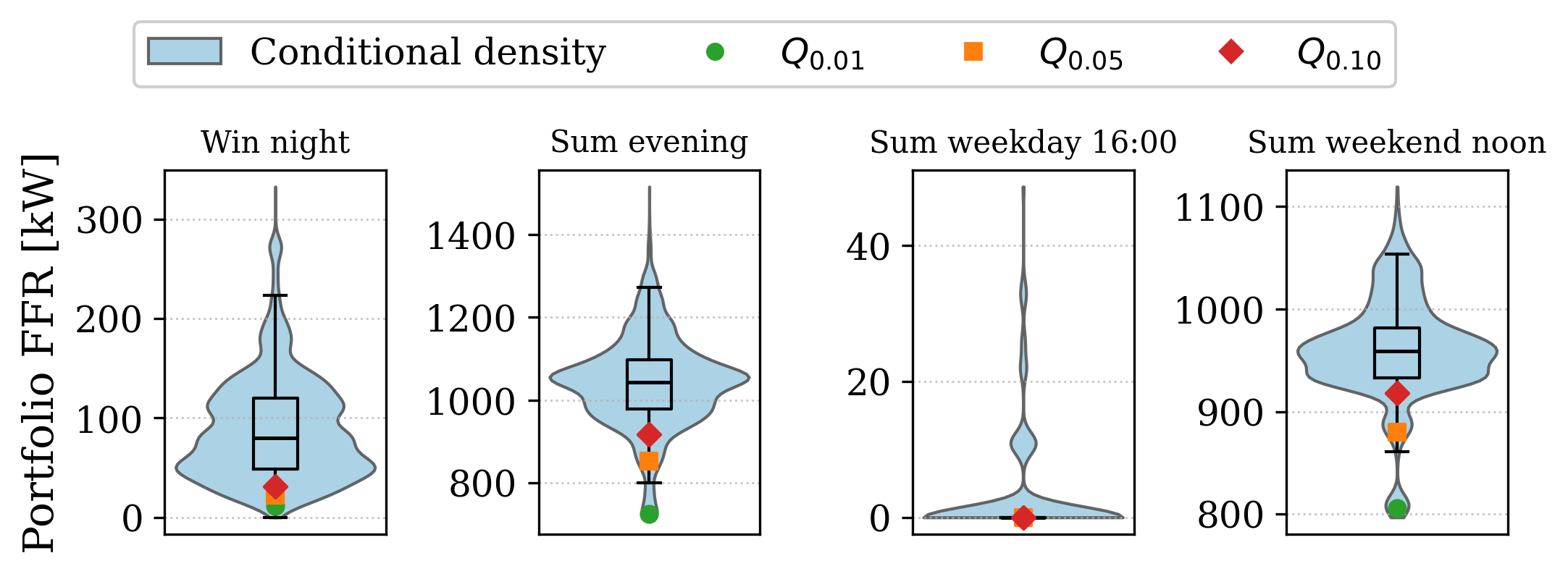}
    \includegraphics[width=0.8\columnwidth]{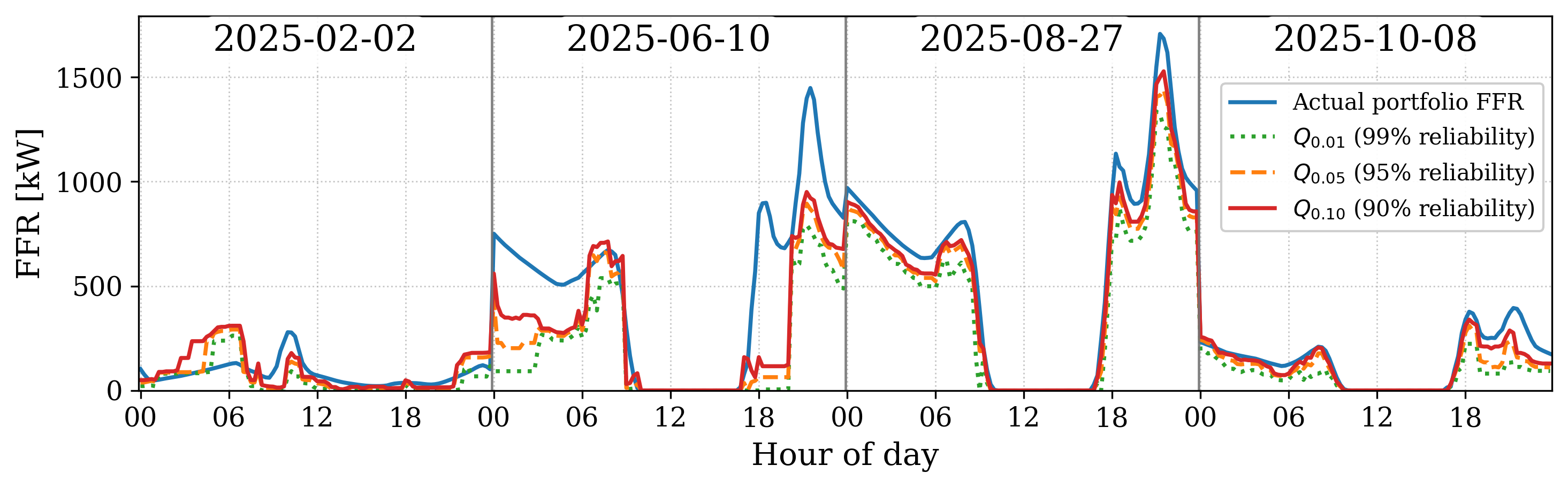}
    \caption{HP QRF results. Top: conditional distributions of portfolio FFR at representative weather--calendar operating points, with $Q_{0.01}$, $Q_{0.05}$, and $Q_{0.10}$ marked. Bottom: day-ahead quantile bids versus realized portfolio FFR on four test days.}
    \label{fig:results_qrf_hp}
\end{figure}

Fig.~\ref{fig:results_qrf_hp} reports the analogous HP results. Fig.~\ref{fig:results_qrf_hp}~(a) shows conditional FFR densities at four weather--calendar operating points, with $Q_{0.01}$, $Q_{0.05}$, and $Q_{0.10}$ marked. Winter night provides moderate heating headroom with a compact lower tail. Summer evening is the highest of the four points, where occupancy and cooling demand overlap. Summer weekend noon retains high cooling-driven capability, below the evening peak. On a summer weekday at 16:00, inside the window in which the schedules keep the heat pumps off, the conditional distribution collapses to a narrow mass at zero and the three bids sit at the origin, so the request in that window can be met only by the EVs. Fig.~\ref{fig:results_qrf_hp}~(b) shows day-ahead quantile bids versus realized HP FFR on four test days: the offers follow the diurnal shape, including the weekday afternoon valley, and stay below the actual for most hours.

Table~\ref{tab:results_qrf_metrics} reports empirical coverage $P(y\ge c)$ and the bid-to-actual ratio $\bar{c}/\bar{y}$ on the test set. Coverage measures how often the realized headroom respects the bid and should match the nominal reliability $1-\varepsilon$. The ratio $\bar{c}/\bar{y}$ summarizes volume efficiency once that reliability is met. Values near unity mean that most of the available headroom is used, whereas low values indicate unused capacity. 
For EVs, coverages of $98.7\%$, $95.5\%$, and $91.5\%$ align closely with the targets $99\%/95\%/90\%$, while $\bar{c}/\bar{y}$ remains high ($0.92$--$0.95$), so the offers are both well calibrated and volume-efficient. For HPs, coverages of $97.8\%$, $94.1\%$, and $90.5\%$ almost meet the same targets, confirming reliable lower-tail bids under the day-ahead feature set of~\eqref{eq:ffr_features}. The $\bar{c}/\bar{y}$ is relatively lower than EVs ($0.58$--$0.77$). The lower $\bar{c}/\bar{y}$ for HPs reflects higher conditional uncertainty; HP FFR coincides with the scheduled load and varies strongly with weather, calendar, and occupancy. EV offers are more volume-efficient because a large share of the headroom comes from V2G discharge, which depends mainly on binary home availability, often near one for residential fleets in overnight/evening windows, while charging load is driven chiefly by weekday/weekend patterns rather than season.

\begin{table}
\centering
\caption{QRF bid performance on the test set.}
\label{tab:results_qrf_metrics}
\begin{tabular}{@{}lccc@{}}
\toprule
Tech. & $\varepsilon$ (Reliability) & Emp.\ coverage & $\bar{c}/\bar{y}$ \\
\midrule
EV & $0.01$ ($99\%$) & $0.987$ & $0.92$ \\
EV & $0.05$ ($95\%$) & $0.955$ & $0.94$ \\
EV & $0.10$ ($90\%$) & $0.915$ & $0.95$ \\
HP & $0.01$ ($99\%$) & $0.978$ & $0.58$ \\
HP & $0.05$ ($95\%$) & $0.941$ & $0.71$ \\
HP & $0.10$ ($90\%$) & $0.905$ & $0.77$ \\
\bottomrule
\end{tabular}
\end{table}

\subsection{Market Clearing and VoLL Comparison}
\label{sec:results_clearing}

\subsubsection{Market Clearing Cost Analysis}
The TSO request $R_t^{\mathrm{FFR}}$ is cleared against the aggregator capacity--price offers using the cost-minimization problem~\eqref{eq:ffr_cost_clearing}. Table~\ref{tab:ffr_clearing_2025} reports annual 2025 costs under the three bid-price scenarios of Table~\ref{tab:ffr_price_scenarios}. For the proposed dynamic request, offers are cleared at $Q_{0.01}$, $Q_{0.05}$, and $Q_{0.10}$ (bid prices held fixed across $Q$, so differences reflect offered volume only). The three non-dynamic benchmarks, seasonal monthly maximum, constant annual 90th percentile, and maximum of the dynamic series, are cleared with the $Q_{0.10}$ offer, matching a less conservative but volume-efficient aggregator posture for the static comparisons.

\begin{table}
\centering
\caption{Annual 2025 FFR clearing cost versus the UFLS VoLL benchmark. Offer quantiles $Q_{0.01}$/$Q_{0.05}$/$Q_{0.10}$ apply to the proposed neighbourhood dynamic request; seasonal and annual benchmarks use $Q_{0.10}$. The bottom panel summarises the real 2025 UFLS events and VoLL cost.}
\label{tab:ffr_clearing_2025}
\label{tab:ufls_voll_2025}
\setlength{\tabcolsep}{2.5pt}
\begin{tabular}{@{}llcccc@{}}
\toprule
Request rule & Offer & Mean $R$ & \multicolumn{3}{c}{Cost [\euro]} \\
\cmidrule(lr){4-6}
 & & [MW] & $S_{-15\%}$ & $S_{\mathrm{central}}$ & $S_{+15\%}$ \\
\midrule
\multirow{3}{*}{Proposed dynamic}
 & $Q_{0.01}$ & 9.20 & 439\,066 & 516\,549 & 594\,031 \\
 & $Q_{0.05}$ & 9.20 & 408\,804 & 480\,945 & 553\,087 \\
 & $Q_{0.10}$ & 9.20 & 396\,447 & 466\,408 & 536\,369 \\
\midrule
Seasonal maximum & $Q_{0.10}$ & 11.46 & 563\,740 & 663\,223 & 762\,707 \\
Annual 90th perc. & $Q_{0.10}$ & 11.53 & 566\,976 & 667\,031 & 767\,086 \\
Annual maximum & $Q_{0.10}$ & 11.80 & 589\,700 & 693\,765 & 797\,830 \\
\midrule
\multicolumn{2}{c}{UFLS VoLL benchmark} & --- & \multicolumn{3}{c}{$637\,500$} \\
\midrule
\multicolumn{2}{@{}c}{Disturbances} & Clear UFLS & \multicolumn{2}{c}{Shed energy} & VoLL \cite{ACER2024VoLL} \\
\multicolumn{2}{@{}c}{screened} & ($f_{\min}\le 49\,\mathrm{Hz}$) & \multicolumn{2}{c}{$E^{\mathrm{shed}}$} & [\euro/MWh] \\
\cmidrule(lr){1-2}\cmidrule(lr){3-3}\cmidrule(lr){4-5}\cmidrule(lr){6-6}
\multicolumn{2}{@{}c}{28} & 15 & \multicolumn{2}{c}{$42.5\,\mathrm{MWh}$} & $15{,}000$ \\
\bottomrule
\end{tabular}
\end{table}

On the dynamic request alone, relaxing the offer from $Q_{0.01}$ to $Q_{0.05}$ and $Q_{0.10}$ lowers cost under central prices by about \euro$36{,}000$ ($\approx 7\%$) and \euro$50{,}000$ ($\approx 10\%$), with a similar relative effect under $S_{\pm 15\%}$. Absolute costs scale nearly linearly with the price factor $\alpha_s$ for the three price scenarios. Under the proposed dynamic request with $Q_{0.10}$ and central prices, the annual cost is approximately \euro$466{,}000$ (mean request $9.20\,\mathrm{MW}$). Relative to the annual maximum, the seasonal maximum, and the annual 90th-percentile constant under central prices, the dynamic $Q_{0.10}$ schedule reduces cost by about \euro$227{,}000$ ($\approx 33\%$), \euro$197{,}000$ ($\approx 30\%$), and \euro$201{,}000$ ($\approx 30\%$), respectively.

\subsubsection{Market Cost and VoLL Comparison}
The annual FFR cost in Table~\ref{tab:ffr_clearing_2025} is benchmarked against the socioeconomic impact of insufficient FFR provision, valued at a VoLL of $15{,}000\,\text{\euro}/\mathrm{MWh}$ as adopted for the European Resource Adequacy Assessment in ACER Decision No.~09/2024~\cite{ACER2024VoLL}. During 2025, 28 frequency disturbances were screened. According to the Cyprus UFLS scheme~\cite{tsoc_operating_margin}, load shedding initiates at $49.0\,\mathrm{Hz}$. As summarised in the lower panel of Table~\ref{tab:ffr_clearing_2025}, fifteen clear UFLS events with $f_{\min}\le 49\,\mathrm{Hz}$ yielded approximately $42.5\,\mathrm{MWh}$ of disconnected energy, comprising net load and distributed PV on the shed feeders, corresponding to $C^{\mathrm{VoLL}}=E^{\mathrm{shed}}\times\mathrm{VoLL}=637{,}500\,\text{\euro}$. Relative to this benchmark, the proposed dynamic request under central prices remains below the VoLL cost for all three offer quantiles and bidding scenarios, whereas the seasonal and annual procurement rules approach or exceed it.

\subsection{Sensitivity Analysis and Disturbance Validation}
\label{sec:results_sensitivity}
\begin{figure}[!ht]
    \centering
    \includegraphics[width=0.8\linewidth]{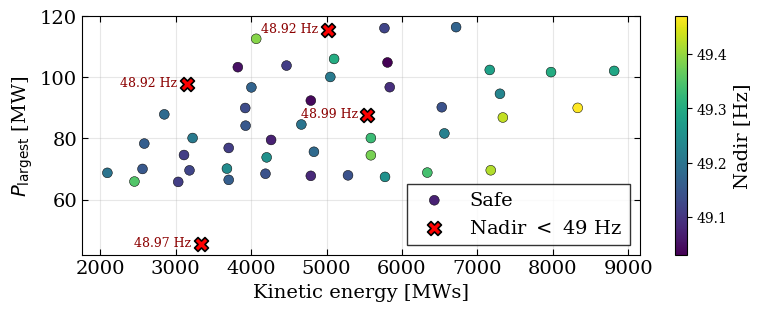}
    \caption{Frequency nadir for the 2025 validation scenarios in the kinetic energy--$P_{\mathrm{largest}}$ plane under 100\% delivery of the cleared FFR. Safe cases are colored by nadir; red markers indicate nadir below 49~Hz.}
    \label{fig:nadir_validation_2025}
\end{figure}

\begin{figure}
    \centering
    \includegraphics[width=0.8\linewidth]{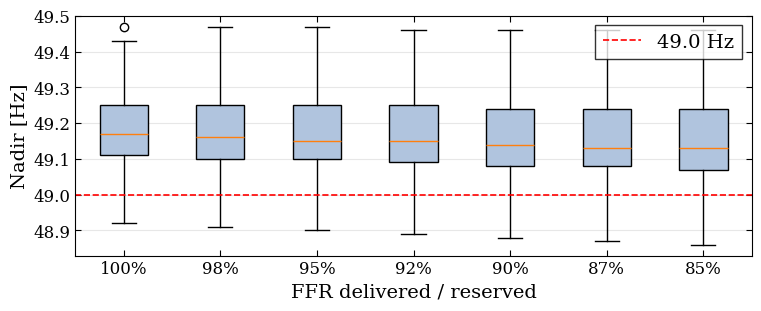}
    \caption{Sensitivity of the frequency nadir to partial FFR delivery (100\%--85\% of the cleared quantity) for the same 2025 validation set. The dashed line marks the 49~Hz UFLS threshold.}
    \label{fig:sensitivity_delivery}
\end{figure}

Similar to the training phase, 50 representative operating points are selected from 2025 for disturbance validation and delivery sensitivity analysis. Under 100\% activation of the cleared FFR, 46 of the 50 points remain above $49\,\mathrm{Hz}$ ($92\%$), as shown in Fig.~\ref{fig:nadir_validation_2025}; the four violations have nadirs between $48.92$ and $48.99\,\mathrm{Hz}$. Fig.~\ref{fig:sensitivity_delivery} shows that at $85\%$ delivery the same four points still violate and no additional UFLS-threshold breaches appear; only the nadirs decrease slightly, so again 46 of 50 points stay above $49\,\mathrm{Hz}$.


\section{Conclusions}
\label{sec:conclusion}

This paper presented an integrated, market-compatible framework for quantifying and procuring Fast Frequency Response in low-inertia systems. On the TSO side, an offline neighbourhood FFR need surface was combined with day-ahead kinetic-energy forecasts to produce a dynamic capacity request for each market interval. On the aggregator side, electric-vehicle and heat-pump fleets were mapped to FFR headroom and offered through risk-aware lower quantiles estimated by quantile regression forests. The resulting capacity--price offers were cleared in a day-ahead pay-as-bid FFR capacity market.

The Cyprus 2025 case study showed that the proposed dynamic request procures about $9.2\,\mathrm{MW}$ on average. Under central prices, this lower mean request reduces the annual cost by about $30\%$--$33\%$ relative to the seasonal and annual static rules. The annual cost lies below the UFLS VoLL of \euro$637{,}500$ estimated from the fifteen real 2025 events with $f_{\min}\le 49\,\mathrm{Hz}$, which provides a reference for whether procuring FFR is worth its cost relative to the conventional practices currently used. Risk-aware EV and HP offers met empirical coverages close to the nominal $99\%/95\%/90\%$ targets, with EVs providing higher volume efficiency than HPs. On 50 representative operating points, 46 remained above $49\,\mathrm{Hz}$ at both $100\%$ and $85\%$ delivery of the cleared FFR. The same separation, a condition-dependent request on the TSO side and a certifiable lower-quantile offer on the aggregator side, applies to other low-inertia systems that already procure FFR as a standalone day-ahead capacity product.

\section*{Acknowledgments} 
This project has received funding from the European Union’s Framework Program for Research and Innovation Horizon Europe under the Marie Skłodowska-Curie Actions Doctoral Networks, Grant Agreement No. 101120278.\\

\bibliographystyle{elsarticle-num} 
\bibliography{ref}

@ARTICLE{10960637,
  author={Lunde, Gustav A. and Damm, Emil V. and Gade, Peter A.V. and Kazempour, Jalal},
  journal={IEEE Transactions on Power Systems}, 
  title={Aggregator of Electric Vehicles Bidding in Nordic FCR-D Markets: A Chance-Constrained Program}, 
  year={2025},
  volume={40},
  number={6},
  pages={4648-4659},
  doi={10.1109/TPWRS.2025.3558868}}

@article{PANAGI2025101699,
title = {Sizing of Fast Frequency Response Reserves for improving frequency security in low-inertia power systems},
journal = {Sustainable Energy, Grids and Networks},
volume = {42},
pages = {101699},
year = {2025},
issn = {2352-4677},
doi = {https://doi.org/10.1016/j.segan.2025.101699},
author = {Savvas Panagi and Petros Aristidou},
}

@techreport{Future-System-Inertia,
  title       = {Future System Inertia 2},
  author = {{ENTSOE}},
  year        = {2017},
    url         = {https://eepublicdownloads.entsoe.eu},
}

@article{iet_savvas_impact_ffr,
author = {Savvas Panagi  and Antonis Lazari  and Vrahimis Koutsoloukas  and Petros Aristidou },
title = {Impact of fast frequency response on renewable energy source curtailment and load shedding},
journal = {IET Conference Proceedings},
volume = {2024},
issue = {5},
pages = {505-508},
year = {2025},
doi = {10.1049/icp.2024.2088},
}

@inproceedings{argyrou2024synergy,
  title={Comparison of fast frequency response methods in the low-inertia grid of cyprus},
  author={Argyrou, Andreas and Panagi, Savvas and Aristidou, Petros},
  booktitle={Conference on Energy Transition in the Mediterranean Area},
  pages={1--5},
  year={2024},
  organization={IEEE}
}

@article{KUSHWAHA2023106944,
title = {Fast frequency response constrained electric vehicle scheduling for low inertia power systems},
journal = {Journal of Energy Storage},
volume = {62},
pages = {106944},
year = {2023},
issn = {2352-152X},
doi = {https://doi.org/10.1016/j.est.2023.106944},
author = {Priyanka Kushwaha and Vivek Prakash and Sumanth Yamujala and Rohit Bhakar},
}

@INPROCEEDINGS{panagioptimal,
  author={Panagi, Savvas and Therapontos, Phivos and Spanias, Chrysovalantis and Aristidou, Petros},
  booktitle={2025 IEEE Kiel PowerTech},
  title={Optimal Operation of Electric Vehicles to Enhance the Flexibility Provision in Active Distribution Grids},
  year={2025},
  doi={10.1109/PowerTech59965.2025.11180664}
}

@inproceedings{panagi2026ifac,
  author={Panagi, Savvas and Spanias, Chrysovalantis and Aristidou, Petros},
  title={Network-Constrained Flexibility Quantification of Heat Pumps with Integrated Space Heating and Storage Tanks in Active Distribution Grids},
  booktitle={IFAC World Congress},
  year={2026},
}

@article{song2024power,
  title={Power-hardware-in-the-loop validation of air-source heat pump for fast frequency response applications},
  author={Song, Ruihao and Mohapatra, Anurag and Hamacher, Thomas and Peri{\'c}, Vedran S},
  journal={EPSR},
  volume={235},
  pages={110754},
  year={2024},
  publisher={Elsevier}
}

@article{song2024integrating,
  title={Integrating air-source heat pumps into the demand-side fast frequency response service: a study based on thermal dynamic uncertainty},
  author={Song, Ruihao and Terzija, Vladimir and Hamacher, Thomas and Peri{\'c}, Vedran S},
  journal={IEEE Transactions on Sustainable Energy},
  volume={16},
  number={1},
  pages={323--335},
  year={2024},
  publisher={IEEE}
}

@article{xu2023quantile,
  title={Quantile regression based probabilistic forecasting of renewable energy generation and building electrical load: A state of the art review},
  author={Xu, Chengliang and Sun, Yongjun and Du, Anran and Gao, Dian-ce},
  journal={Journal of Building Engineering},
  volume={79},
  pages={107772},
  year={2023},
  doi={10.1016/j.jobe.2023.107772}
}

@article{zhang2023xgboost,
  title={XGBoost-based short-term prediction method for power system inertia and its interpretability},
  author={Zhang, Lei and Guo, Zhihao and Tao, Qianhui and Xiong, Zhizhi and Ye, Jing},
  journal={Energy Reports},
  volume={9},
  pages={1458--1469},
  year={2023},
  publisher={Elsevier}
}

@article{meinshausen2006quantile,
  title={Quantile regression forests.},
  author={Meinshausen, Nicolai and Ridgeway, Greg},
  journal={Journal of machine learning research},
  volume={7},
  number={6},
  year={2006}
}

@dataset{dimonti_2026_19364191,
  author       = {Dimonti, Noah},
  title        = {EV Charging Optimisation Dataset},
  month        = apr,
  year         = 2026,
  publisher    = {Zenodo},
  doi          = {10.5281/zenodo.19364191},
  url          = {https://doi.org/10.5281/zenodo.19364191},
}

@techreport{IEA2023NetZero,
  author      = {{International Energy Agency}},
  title       = {Net Zero Roadmap: A Global Pathway to Keep the 1.5~{\textdegree}C Goal in Reach},
  address     = {Paris, France},
  year        = {2023},
}

@article{VATSAL2025111778,
title = {Fast frequency response from electric buses in low inertia grids},
journal = {Electric Power Systems Research},
volume = {247},
pages = {111778},
year = {2025},
issn = {0378-7796},
author = {Deven Vatsal and Kailash Chand Sharma and Vivek Sharma},
}

@ARTICLE{7115982,
  author={Teng, Fei and Trovato, Vincenzo and Strbac, Goran},
  journal={IEEE Transactions on Power Systems}, 
  title={Stochastic Scheduling With Inertia-Dependent Fast Frequency Response Requirements}, 
  year={2016},
  volume={31},
  number={2},
  pages={1557-1566},
  doi={10.1109/TPWRS.2015.2434837}}

@ARTICLE{9470941,
  author={Garcia, Manuel and Baldick, Ross},
  journal={IEEE Transactions on Power Systems}, 
  title={Requirements for Interdependent Reserve Types Providing Primary Frequency Control}, 
  year={2022},
  volume={37},
  number={1},
  pages={51-64},
  doi={10.1109/TPWRS.2021.3094214}}

@article{DEVOS2019272,
title = {Dynamic dimensioning approach for operating reserves: Proof of concept in Belgium},
journal = {Energy Policy},
volume = {124},
pages = {272-285},
year = {2019},
issn = {0301-4215},
author = {K. {De Vos} and N. Stevens and O. Devolder and A. Papavasiliou and B. Hebb and J. Matthys-Donnadieu},
}

@misc{fingrid_ffr,
  author       = {{Fingrid Oyj}},
  title        = {Fast Frequency Reserve},
  howpublished = {\url{https://www.fingrid.fi/en/electricity-market/reserves/reserve-products/fast-frequency-reserve-ffr/}},
  note         = {Accessed: Sep. 5, 2026}
}

@misc{statnett_ffr,
  author       = {{Statnett}},
  title        = {Raske frekvensreserver -- FFR},
  howpublished = {\url{https://www.statnett.no/for-aktorer-i-kraftbransjen/reservemarkeder/ffr/}},
  note         = {Accessed: Sep. 5, 2026}
}

@misc{tsoc_operating_margin,
  author       = {{Transmission System Operator Cyprus}},
  title        = {Operating Margin Policy},
  howpublished = {\url{https://tsoc.org.cy/electrical-system/operating-margin-policy/}},
  note         = {Accessed: Sep. 25, 2026}
}

@article{crawley2001energyplus,
  title={EnergyPlus: creating a new-generation building energy simulation program},
  author={Crawley, Drury B and Lawrie, Linda K and Winkelmann, Frederick C and Buhl, Walter F and Huang, Y Joe and Pedersen, Curtis O and Strand, Richard K and Liesen, Richard J and Fisher, Daniel E and Witte, Michael J and others},
  journal={Energy and buildings},
  volume={33},
  number={4},
  pages={319--331},
  year={2001},
  publisher={Elsevier}
}

@inproceedings{kourougianni2024strategies,
  title={Strategies for optimizing energy efficiency in residential buildings: an integrated framework approach},
  author={Kourougianni, Fanourios and Olympios, Andreas V and Arsalis, Alexandros and Georghiou, George E},
  booktitle={Conference on Energy Transition in the Mediterranean Area (SyNERGY MED)},
  pages={1--5},
  year={2024},
  organization={IEEE}
}

@misc{SPSL_TSOC_Data_Analysis,
  author       = {{Sustainable Power Systems Lab (SPSL)}},
  title        = {{TSOC Data Analysis}},
  howpublished = {\url{https://tsoc-data-analysis.sps-lab.org/}},
  note         = {Software documentation, accessed: 28 September 2026}
}

@techreport{ENTSOE_CE_SAFA_2024,
  author      = {{ENTSO-E}},
  title       = {Policy on Load-Frequency Control and Reserves},
  year        = {2024},
  url         = {https://eepublicdownloads.entsoe.eu/clean-documents/SOC%20documents/SAFA_for_RG_CE/SAFA_for_RG_CE-03-Annex_01-Policy_on_Load-Frequency_Control_and_Reserves_240724.pdf},
  note        = {Accessed: 2026-09-29}
}

@techreport{NordicSOA_FCR_2025,
  author      = {{Nordic Transmission System Operators}},
  title       = {Appendix 1 to Annex LFCR},
  year        = {2025},
  url         = {https://www.fingrid.fi/globalassets/dokumentit/fi/kantaverkko/kantaverkon-kehittaminen/osa-pohjoismaista-sahkojarjestelmaa/nordic-soa_appendix-1-to-annex-lfcr_regularly-changing-parameters_2025.pdf},
  note        = {Accessed: 2026-09-29}
}

@techreport{ACER2024VoLL,
  author      = {{European Union Agency for the Cooperation
                  of Energy Regulators (ACER)}},
  title       = {{Decision No. 09/2024}},
  year        = {2024},
  month       = jul,
  day         = {5},
  number      = {09/2024},
  type        = {Decision},
  url         = {https://eepublicdownloads.entsoe.eu/clean-documents/nc-tasks/Decision%20No%2009-2024.pdf},
  note        = {Section 6.2.2, paragraphs (43) and (46), pp. 8--10}
}

@dataset{panagi_2026_23556477,
  author       = {Panagi, Savvas},
  title        = {Risk-Aware FFR Quantification Dataset - EVs/HPs},
  month        = oct,
  year         = 2026,
  publisher    = {Zenodo},
  doi          = {10.5281/zenodo.23556477},
  url          = {https://doi.org/10.5281/zenodo.23556477},
}

@article{manner2024ev_ffr,
author = {Pekka Manner and Ville Tikka and Samuli Honkapuro and Kyosti Tikkanen and Jamshid Aghaei},
title = {Electric vehicle charging as a source of {N}ordic fast frequency reserve: proof of concept},
journal = {IET Generation, Transmission \& Distribution},
volume = {18},
number = {6},
pages = {1223-1234},
year = {2024},
doi = {10.1049/gtd2.13042},
}

@techreport{NordicTSO2019FFR,
  author      = {{Energinet and Fingrid and Statnett and Svenska kraftn\"at}},
  title       = {Fast Frequency Reserve: Solution to the {N}ordic Inertia Challenge},
  institution = {Nordic TSOs},
  year        = {2019},
  month       = dec,
  url         = {https://www.statnett.no/globalassets/for-aktorer-i-kraftsystemet/utvikling-av-kraftsystemet/nordisk-frekvensstabilitet/ffr-stakeholder-report_13122019.pdf},
}

@techreport{NAG2025MinInertia,
  author      = {{Nordic Analysis Group}},
  title       = {Requirements for Minimum Inertia in the {N}ordic Power System},
  institution = {Energinet, Fingrid, Statnett and Svenska kraftn\"at},
  year        = {2025},
  month       = aug,
  url         = {https://eepublicdownloads.blob.core.windows.net/public-cdn-container/clean-documents/SOC%20documents/Nordic/2025/2025_-_Requirements_for_minimum_inertia_in_the_Nordic_power_system.pdf},
}

@misc{NESO2020DC,
  title        = {National {G}rid {ESO} debuts {D}ynamic {C}ontainment frequency response service},
  year         = {2020},
  month        = sep,
}

@article{heylen2022inertia,
author = {Evelyn Heylen and Jethro Browell and Fei Teng},
title = {Probabilistic Day-Ahead Inertia Forecasting},
journal = {IEEE Transactions on Power Systems},
volume = {37},
number = {5},
pages = {3738-3746},
year = {2022},
doi = {10.1109/TPWRS.2021.3134811},
}

@article{riquelme2024ke,
author = {Jose Miguel Riquelme-Dominguez and Manuel Carranza-Garc\'ia and Pedro Lara-Ben\'itez and Francisco M. Gonzalez-Longatt},
title = {A machine learning-based methodology for short-term kinetic energy forecasting with real-time application: {N}ordic Power System case},
journal = {International Journal of Electrical Power \& Energy Systems},
volume = {156},
pages = {109730},
year = {2024},
doi = {10.1016/j.ijepes.2023.109730},
}

@article{Callaway2009,
  author  = {Callaway, Duncan S.},
  title   = {Tapping the Energy Storage Potential in Electric
             Loads to Deliver Load Following and Regulation,
             with Application to Wind Energy},
  journal = {Energy Conversion and Management},
  volume  = {50},
  number  = {5},
  pages   = {1389--1400},
  year    = {2009},
  doi     = {10.1016/j.enconman.2008.12.012},
  url     = {https://doi.org/10.1016/j.enconman.2008.12.012}
}

@article{Nema2025,
  author  = {Nema, Sumit and Prakash, Vivek
             and Pand{\v{z}}i{\'c}, Hrvoje},
  title   = {The Role of Thermostatically Controlled Loads
             in Power System Frequency Management:
             A Comprehensive Review},
  journal = {Sustainable Energy, Grids and Networks},
  volume  = {42},
  pages   = {101680},
  year    = {2025},
  doi     = {10.1016/j.segan.2025.101680},
  url     = {https://doi.org/10.1016/j.segan.2025.101680}
}

@misc{FujitsuKMCE2022,
  author       = {{Fujitsu General Limited}},
  title        = {Design and Technical Manual for Wall-Mounted
                  Inverter Air Conditioners},
  year         = {2022},
  howpublished = {Manufacturer Technical Documentation},
  note         = {Models ASYG07KMCE, ASYG09KMCE,
                  ASYG12KMCE and ASYG14KMCE.
                  Document DR\_AS158EF\_01},
  url          = {https://webstore.uk.fujitsu-general.com/product/attachment/ASYG09KMCE/design%20and%20technical%20manual.pdf}
}

@manual{DaikinFTXMA2024,
  author       = {{Daikin}},
  title        = {FTXM-A Wall Mounted Unit:
                  Technical Data},
  organization = {Daikin},
  year         = {2024},
  note         = {Technical Data Book EEDEN24.
                  Residential wall-mounted
                  inverter split systems},
  url          = {https://my.daikin.eu/content/dam/b2c/south-africa/dasa-pdf/new/FTXM-A_Data%20book_EEDEN24A_English.pdf/_jcr_content/renditions/original./FTXM-A_Data%20book_EEDEN24A_English.pdf}
}

@article{Crawley2023,
  author  = {Crawley, Jenny and Martin-Vilaseca, Adria
             and Wingfield, Jez and Gill, Zachary
             and Shipworth, Michelle and Elwell, Clifford},
  title   = {Demand Response with Heat Pumps: Practical
             Implementation of Three Different Control Options},
  journal = {Building Services Engineering Research and Technology},
  volume  = {44},
  number  = {2},
  pages   = {211--228},
  year    = {2023},
  doi     = {10.1177/01436244221145871},
  url     = {https://doi.org/10.1177/01436244221145871}
}

@article{Kim2016,
  author  = {Kim, Young-Jin and Fuentes, Emmanuel
             and Norford, Leslie K.},
  title   = {Experimental Study of Grid Frequency Regulation
             Ancillary Service of a Variable Speed Heat Pump},
  journal = {IEEE Transactions on Power Systems},
  volume  = {31},
  number  = {4},
  pages   = {3090--3099},
  year    = {2016},
  doi     = {10.1109/TPWRS.2015.2472497},
  url     = {https://doi.org/10.1109/TPWRS.2015.2472497}
}

@techreport{cystat_housing,
  author      = {{Statistical Service of Cyprus}},
  title       = {Construction and Housing Statistics 2011},
  institution = {Statistical Service of Cyprus},
  address     = {Nicosia},
  year        = {2013},
  note        = {Series II, Report No. 34}
}

@techreport{cyprus_costoptimal2023,
  author      = {{Frederick University}},
  title       = {Calculation of the cost-optimal levels of minimum energy performance requirements for buildings in accordance with {Article} 5 of {Directive} 2010/31/{EU}},
  institution = {Energy Service, Ministry of Energy, Commerce and Industry of Cyprus},
  address     = {Nicosia, Cyprus},
  year        = {2023},
  note        = {Final Report, contract YEEB/YE/03/2022, 31 July 2023, rev. 7 Feb. 2024 (in Greek)}
}

@misc{ncm_activity_db,
  author       = {{Building Research Establishment (BRE)}},
  title        = {National Calculation Methodology ({NCM}) activity database, public release of 15 December 2021},
  howpublished = {\url{https://www.uk-ncm.org.uk}},
  year         = {2021}
}

@article{olympios2025stochastic,
  author  = {Olympios, Andreas V. and Mersch, Matthias and Kourougianni, Fanourios and Markides, Christos N. and Pantaleo, Antonio M. and Kyprianou, Andreas and Georghiou, George E.},
  title   = {A stochastic optimisation framework for integrating photovoltaic systems, heat pumps, and energy storage in buildings},
  journal = {Applied Thermal Engineering},
  volume  = {278},
  pages   = {127312},
  year    = {2025},
  doi     = {10.1016/j.applthermaleng.2025.127312}
}

@article{hersbach2020era5,
  author  = {Hersbach, Hans and Bell, Bill and Berrisford, Paul and Hirahara, Shoji and Hor{\'a}nyi, Andr{\'a}s and Mu{\~n}oz-Sabater, Joaqu{\'i}n and others},
  title   = {The {ERA5} global reanalysis},
  journal = {Quarterly Journal of the Royal Meteorological Society},
  volume  = {146},
  number  = {730},
  pages   = {1999--2049},
  year    = {2020},
  doi     = {10.1002/qj.3803}
}

@misc{zippenfenig2023openmeteo,
  author    = {Zippenfenig, Patrick},
  title     = {Open-Meteo.com Weather {API}},
  year      = {2023},
  publisher = {Zenodo},
  doi       = {10.5281/zenodo.7970649}
}
\end{document}